\documentclass[amsmath,eqsecnum,preprint,pre,aps,showpacs]{revtex4}
\usepackage{amsmath}
\usepackage{graphicx}
\usepackage{dcolumn}
\usepackage{bm}
\usepackage{float}
\usepackage{xcolor}

\begin{document}

\draft
\emph{}
\title{From equilibrium to non-equilibrium \\ statistical mechanics of liquids.}
\author{O. Joaqu\'in-Jaime$^{1}$,  R. Peredo-Ortiz$^{1,2}$,  M. Medina-Noyola$^{1}$ and L.F. Elizondo-Aguilera$^{4}$.}
\affiliation{$^{1}$ Instituto de F\'{\i}sica,
Universidad Aut\'{o}noma de San Luis Potos\'{\i}, \'{A}lvaro
Obreg\'{o}n 64, 78000 San Luis Potos\'{\i}, SLP, M\'{e}xico}
\affiliation{$^{2}$ Facultad de Ciencias F\'isico-Matem\'aticas,
Benem\'erita Universidad Aut\'{o}noma de Puebla, Apartado Postal 
1152, CP 72570, Puebla, PUE, M\'{e}xico}
\affiliation{$^4$ Instituto de F\'isica, Benem\'erita Universidad Aut\'onoma de Puebla, 
Apartado Postal J-48, 72570 Puebla, Mexico.}

\date{\today}

\begin{abstract}

Relevant and fundamental concepts of the statistical mechanical theory of classical liquids are ordinarily introduced in the context of the description of thermodynamic equilibrium states. This makes explicit reference to probability distribution functions of \emph{equilibrium} statistical ensembles (canonical, microcanonical, ...) in the derivation of general and fundamental relations between inter-particle interactions and measurable macroscopic properties of a given system. This includes, for instance, expressing the internal energy and the pressure as functionals of the radial distribution function, or writing transport coefficients (diffusion constant, linear viscosity, ...) in terms of integral relations involving both, static and dynamic auto-correlation functions (density-density, stress-stress, ...). Most commonly, however, matter is not in thermodynamic equilibrium, and this calls for the extension of these relations to out-of-equilibrium conditions with the aim of understanding, for example, the time-dependent transient states during the process of equilibration, or the aging of glass- and gel-forming liquids during the formation of non-equilibrium amorphous solid states. In this work we address this issue from both, a general perspective and an illustrative concrete application focused on the first principles description of rheological and viscoelastic properties of glass- and gel-forming liquids.

\end{abstract}


\maketitle

\section{Introduction}\label{section1}

The thermodynamic and statistical mechanical description of equilibrium liquids rests on firm and well-established fundamental basis \cite{frischlebowitz}. Thermodynamic concepts such as equations of state, equilibrium phases and phase diagrams \cite{callen}, as well as statistical mechanical concepts such as pair correlation functions, the Ornstein-Zernike equation, or free energy density functionals, are nowadays well-understood textbook material  \cite{evans,mcquarrie,hansen,boonyip}. One reason for the beauty and simplicity of these concepts is that their ordinary definition and application only refers to macroscopic states contained in the universal set (or ``catalog'') of thermodynamic equilibrium states of matter, which are understood in terms of the maximum-entropy principle \cite{callen} together with Boltzmann's expression $S=k_B\ln W$ for the entropy $S$ in terms of the number $W$ of microscopic states \cite{mcquarrie}. In general, when gases, liquids and crystalline solids reach a thermodynamic equilibrium state, their properties are stationary, independent of the preparation protocol, and determined by the solution of the equation $dS[\mathbf{A}]=0$, where $S$ is the total entropy (including reservoirs) and the components of the vector $\mathbf{A}$ are the extensive thermodynamic variables.

In contrast, it is not clear how Boltzmann's principle operates in general to describe, for example, the formation of very common \emph{non-equilibrium} amorphous materials (glasses, gels, etc.), whose properties may exhibit aging and might depend on their preparation  protocol \cite{angellreview1,ngaireview1}.  Although the amorphous solidification of glass- and gel-forming liquids is an ubiquitous non-equilibrium process of enormous relevance in physics, chemistry, biology, and materials science and engineering \cite{dawsonreview}, their  fundamental understanding  is sometimes referred to as \emph{``the deepest and most interesting unsolved problem in solid state theory''} \cite{anderson}. The macroscopic states in which all the known (and unknown) non-equilibrium amorphous materials are found, constitute a second universal catalog of states of matter, additional and disjoint to the catalog of stationary equilibrium states that solve the equation $dS[\mathbf{A}]=0$. 
Thus, one might claim that the referred fundamental problem cannot be declared as ``solved'' until a general and fundamental physical principle is identified, from which one can derive the equation whose solutions, in principle, describe and predict the properties of materials in this second universal catalog of states of matter.  

This fundamental problem, however, is only one of the many concerns of non-equilibrium statistical mechanics, which comprises many different theoretical tools, such as Boltzmann kinetic equation, time correlation function formalism, projection operator techniques,  stochastic equations, the mode-coupling theory, and the dynamic density functional theory, amply described in authoritative textbooks and reviews \cite{nicolisprigogine,zwanzig,casasvazquez0,bagchi,oettinger,tebrugtlowen}, and in the references therein. Somehow, however, and in spite of the long-standing scientific interest and extensive research efforts to extend statistical mechanical methods to non-equilibrium conditions, the progress in its application to the fundamental theoretical understanding of the amorphous solidification of supercooled liquids has been rather modest. 

Let us recall as a reference that almost two centuries ago, Clapeyron summarized the experimental results of Boyle, Charles and Avogadro into the empirical ideal gas equation of state. The need to explain this experimental phenomenology in molecular terms, in turn, led Clausius, Maxwell and Boltzmann to elaborate the kinetic molecular theory of ideal gases, thus inaugurating the theoretical methods of statistical mechanics \cite{schmitzphyschembook}. We may say that the current understanding of non-equilibrium amorphous solids is still in the stage of gathering empirical experimental information, with increasingly greater (even microscopic) detail, particularly when complemented with molecular simulation methods. Although no analogous simple phenomenological synthesis has emerged from the overwhelmingly varied accumulation of experimental data describing all the features that characterize the real physical behavior of glass- and gel-forming liquids  \cite{angellreview1,ngaireview1}, insightful phenomenological models exist that describe relevant features of glass behavior. This is illustrated, for example, by the Tool-Narayanaswamy-Moynihan \cite{tool,narayanaswamy,moynihan1} and the Kovacs-Aklonis-Hutchinson-Ramos \cite{kahr} models, commonly used in industry to predict aging effects \cite{hornboll}, and whose development involved a rich discussion of many relevant issues \cite{hodge}. These phenomenological models intelligently compile many previous partial discoveries, just like Clapeyron compiled the empirical data represented by the ideal gas equation of state $pV=nRT$. 

Continuing with the previous analogy, the notorious missing piece is the first-principles theoretical description of the experimental phenomenology of glass-forming liquids during the process of amorphous solidification. In spite of a rich and well-documented theoretical discussion of relevant aspects of the behavior of  viscous liquids \cite{mauroallan,berthierreview,berthierbook2011,charbonneauparisizamponi}, we are still missing the non-equilibrium analog for glasses, of the molecular statistical mechanical theory developed by Clausius, Maxwell, and Boltzmann to understand ideal gases, which was later extended to equilibrium non-ideal gases by van der Waals \cite{vdw} and, eventually, to liquids by Ornstein and Zernike  \cite{oz1914}, Widom \cite{widom},  and many others (see \cite{evans,mcquarrie,hansen,boonyip}). Building the analog of these developments in the context of non-equilibrium glass- and gel-forming liquids, poses a relevant challenge to the \emph{``beautiful and profound subject''} \cite{bagchi} of non-equilibrium statistical mechanics, which thus has the opportunity to become the theoretical counterpart of experiments and simulations, in the search for the fundamental understanding of non-equilibrium states of matter. Thus, a relevant initiative is now to focus this rich theoretical infrastructure on the specific theoretical challenge of understanding amorphous materials from first principles, including the behavior of liquids during the irreversible transient process of dynamic-arrest (or ``aging''), occurring in highly viscous liquids during their amorphous solidification into glassy and gelled states.

This was precisely the main aim of the recently-developed statistical physics formalism referred to as the \emph{non-equilibrium self-consistent generalized Langevin equation (NE-SCGLE)} theory  \cite{nescgle0,nescgle1,nescgle2,nescgle3,nescgle5}, whose essence is a set of time-evolution equations for the structural and dynamical properties of a non-equilibrium liquid, namely, Eqs. (4.1)-(4.7) of Ref. \cite{nescgle1}. The NE-SCGLE theory  originated, somewhat off the beaten path, from the assumption that the manner in which Boltzmann's postulate $S=k_B\ln W$ explains non-equilibrium states, is provided by a spatially non-local and temporally  non-Markovian and non-stationary generalization \cite{nescgle0} of Onsager's linear irreversible thermodynamics  \cite{onsager1, onsager2} and the Onsager-Machlup theory of thermal fluctuations \cite{onsagermachlup1, onsagermachlup2}. We can say that  the resulting NE-SCGLE equations, and the remarkable predicted scenario they have unveiled, constitute a highly relevant contribution to the foundations of the non-equilibrium extension of the statistical mechanical theory of equilibrium liquids. 

A brief and updated account of the fundamental origins of the NE-SCGLE theory is provided by Ref. \cite{nonlinonsmach} and, hence, here we do not dwell on this subject. Similarly, here we shall not review the applications of the NE-SCGLE theory, i.e., the solution of Eqs. (4.1)-(4.7) of Ref. \cite{nescgle1} for a variety of model systems, which illustrate the competition between the  kinetic  processes of thermodynamic equilibration, and the ultra-slow kinetic  processes of formation of non-equilibrium amorphous solids; these contributions will be the subject of a forthcoming publication \cite{reviewapplications}. Instead, the general aim of the present work is to illustrate in detail the possible strategies to identify the non-equilibrium analog of some well-established elementary concepts of the statistical mechanics of equilibrium liquids. For this we first discuss -- with the support of the NE-SCGLE theory -- the non-equilibrium role of two well-established notions of the (equilibrium) liquid state theory, namely, the Ornstein-Zernike equation and the Wertheim-Lovett relation \cite{evans,mcquarrie,hansen}. 

We then follow a similar route to provide an approximate expression for a highly relevant dynamic property, namely, the frequency-dependent dynamic shear viscosity $\eta(\omega)$, written in terms of the structure factor and intermediate scattering functions. Such an expression was first  derived by Geszti  for atomic fluids \cite{geszti} and  by  N\"agele and Bergenholtz \cite{naegele,banchio} for colloids, albeit only for thermodynamic equilibrium conditions. Here, instead, we shall zoom in on the detailed theoretical arguments leading to a general expression for the non-equilibrium dynamic shear viscosity $\eta(\omega;t)$ of a liquid at a (waiting) time $t$ after being suddenly quenched to arbitrary final temperature and density. This derivation follows a simple strategy, consisting of inspecting the derivation of the equilibrium counterpart, to see if in reality the assumption of equilibrium conditions was really essential, for example, if at some point an explicit and indispensable use was made of any equilibrium statistical ensemble. 

We start this work  in Section \ref{section2} by illustrating this strategy with a simple example, namely, the extension to non-equilibrium of the so called energy equation, which is shown to be identical to the usual expression for the internal energy in terms of the radial distribution function $g(r)$, except that  $g(r)$ is replaced by the $t$-dependent non-equilibrium radial distribution function $g(r;t)$. This poses the crucial questions of how to determine $g(r;t)$.  At equilibrium, $g(r;t)$ is  independent of $t$, and is related with the so-called direct correlation function $c(r)$ by means of the Ornstein-Zernike equation, whose validity or extension at non-equilibrium conditions is also a natural and intriguing question.  In Section \ref{section2} we also show that the NE-SCGLE theory actually provides a straightforward response to both of these issues.

Section \ref{section3} explores a possible route to extend to no-equilibrium conditions another fundamental relation derived and employed in the statistical thermodynamic theory of inhomogeneous fluids at equilibrium, namely, the so called Wertheim-Lovett relation \cite{evans}. This relation writes the gradient of the equilibrium local particle number density $n(\mathbf{r})$ as a convolution of the two-particle correlation function and the pairwise force between particles. From the conventional arguments employed in its equilibrium derivation, it would be understood to be valid only at equilibrium. In Section \ref{section3}, however,  we demonstrate that a particular case of the Wertheim-Lovett relation derives from symmetry considerations that are completely transportable to non-equilibrium conditions.

Section \ref{section4} then focuses on the most ambitious objective of this contribution, namely,the 
derivation of the non-equilibrium expression for the dynamic shear viscosity $\eta(\omega;t)$, which turns out to be almost identical to its equilibrium counterpart. Our present derivation, however, does not really follow in detail the arguments and steps employed in the original derivations, constrained to thermodynamic equilibrium \cite{geszti,naegele,banchio}. However, it is not fundamentally different, except for the fact that our derivation assumes the condition of stationarity, rather than  the condition of thermodynamic equilibrium. To the best of our knowledge, such a closed equation for $\eta(\omega;t)$ has never been proposed before. Finally, in Section \ref{section5} we provide a brief discussion of perspectives and a summary of  conclusions.

\section{Non-equilibrium version of elementary equilibrium concepts} \label{section2}   

The statistical mechanical theory of classical fluids was the subject of active development in the second half of the last century. This development mostly focused on the description of the properties of systems in thermodynamic equilibrium states, as recorded in influential monographs and textbooks \cite{callen,frischlebowitz,evans,mcquarrie,hansen,boonyip}. One of the main aims, for example, was to relate inter-particle interactions with measurable macroscopic properties of a given system, as illustrated, for instance, by the so-called energy equation \cite{mcquarrie,hansen}
\begin{equation}
\frac{U}{N}  =\frac{3}{2}k_BT+ \frac{n}{2}\int   u(r)  g(r)   d^3 r,
\label{energyeq}
\end{equation} 
which expresses the internal energy $U$ of a fluid of $N$ spherical particles in a volume $V$ at temperature $T$ and number density $n=N/V$, in terms of the pair potential $u(r)$ and of the radial distribution function $g(r)$. Similar expressions  were derived  for other thermodynamic properties (e.g. pressure, isothermal compressibility) \cite{mcquarrie,hansen}.  In addition, transport coefficients (e.g. diffusion constant, linear viscosity) were  written in terms of integral relations involving both, structural and dynamical auto-correlation functions (density-density, stress-stress, etc)  \cite{boonyip}, which approximate theories \cite{goetze1,goetze4,goetzebook,hessklein} were able to write in terms of  $u(r)$ and $g(r)$. This, in fact, is another reason why much of the early liquid state theory was centered on the determination of $g(r)$. 

The standard derivation of general equilibrium relations, such as the energy equation above, makes explicit use of probability distribution functions of \emph{equilibrium}  (canonical, microcanonical, ...) statistical ensembles \cite{evans,mcquarrie,hansen,boonyip}. We may thus be conditioned to believe that their validity is restricted to systems in thermodynamic equilibrium. There are, however, many reasons to revise these relations, concepts, and derivations, with the aim of extending them to more general out-of-equilibrium conditions, and here we start precisely with the energy equation.

\subsection{Macroscopic properties and statistical ensembles of non-equilibrium liquids} \label{subsection2.1}

Let us start  by recalling that the microscopic dynamics of a many-body system is governed by the fundamental dynamical (Newton's or Hamilton's) equations describing the motion of each of the $N$ particles comprising the system. Thus, if $\mathbf{r}_i(t)$ denotes the position of the $i$th particle at time $t$ and $\mathbf{p}_i(t)$ its momentum, then the time-evolution of any dynamical variable $\hat A(t)\equiv \hat A(\mathbf{r}^N(t),\mathbf{p}^N(t))$, with  $\mathbf{r}^N(t) \equiv (\mathbf{r}_1(t),\mathbf{r}_2(t), ...,\mathbf{r}_N(t))$ and $\mathbf{p}^N(t) \equiv (\mathbf{p}_1(t),\mathbf{p}_2(t), ...,\mathbf{p}_N(t))$, will be rooted in these microscopic equations of motion, as described in any reference textbook of classical mechanics  \cite{goldstein}. The fundamental postulate of statistical mechanics \cite{mcquarrie,hansen} is that any measurable observable $A(t)$ of a macroscopic system corresponds to the average value of a specific dynamic variable $\hat A\equiv \hat A(\mathbf{r}^N,\mathbf{p}^N)$, i.e., 
\begin{equation}
A(t)=\langle \hat A(t) \rangle \equiv  \int \hat A({\bf r}^N, {\bf p}^N)\ \mathcal{P}_N({\bf r}^N, {\bf p}^N; t)\ d{\bf r}^N\ d{\bf p}^N,
\label{meanvalueofdynvar}
\end{equation}
where the brackets $\langle \cdot \cdot \cdot \rangle$ indicate average over a statistical ensemble, written here in terms of the $N$-particle  probability distribution function (PDF) $\mathcal{P}_N({\bf r}^N, {\bf p}^N; t)$ that represents the conditions imposed on the system. 

Restricting ourselves to thermodynamic equilibrium states (which are strictly stationary) $A(t)=A = \langle\hat A\rangle^{eq}$, where the label $``eq"$ indicates any of the conventional equilibrium  statistical ensembles (canonical, microcanonical, etc.). For instance, in the canonical ensemble we may write $A$ as 
\begin{equation}
A=\langle\hat A\rangle^{eq} \equiv \int \hat A(\mathbf{r}^N,\mathbf{p}^N) \mathcal{P}_N^{eq}(\mathbf{r}^N,\mathbf{p}^N)d \mathbf{r}^N d \mathbf{p}^N, 
\label{meaneq}
\end{equation}
where $\mathcal{P}_N^{eq}(\mathbf{r}^N,\mathbf{p}^N)$ is the equilibrium  $N$-particle PDF, given by 
\begin{equation}
\mathcal{P}_N^{eq}(\mathbf{r}^N,\mathbf{p}^N)\equiv \frac{1}{h^{3N}N!} \frac{e^{-\beta \mathcal{H}(\mathbf{r}^N,\mathbf{p}^N)}}{Q_N}, 
\label{canonicalpdf}
\end{equation}
where $\beta^{-1}=k_BT$, and with $\mathcal{H}(\mathbf{r}^N,\mathbf{p}^N)$ being the Hamiltonian of the system and $Q_N$ the canonical partition function 
\begin{equation}
Q_N\equiv \frac{1}{h^{3N}N!} \int  e^{-\beta \mathcal{H}(\mathbf{r}^N,\mathbf{p}^N)}\ d \mathbf{r}^N d \mathbf{p}^N. 
\label{canonicalpfq}
\end{equation}

Eqs. (\ref{meaneq})-(\ref{canonicalpfq}) provide the fundamental basis for the conventional  statistical mechanical  derivation of general expressions for the thermodynamic observables. For example, Refs. \cite{mcquarrie,hansen} describe in detail the steps and arguments that lead, from these equations, to the expression for the internal energy in  Eq. (\ref{energyeq}). To start with a simple illustrative example, let us now discuss to what extent those arguments and steps can be extended to non-equilibrium conditions.

\subsection{The non-equilibrium energy equation.} \label{subsection2.2}   

Under general non-equilibrium conditions, the macroscopic state of a system may be described by a statistical ensemble, now represented by the time-dependent PDF $\mathcal{P}_N(\mathbf{r}^N, \mathbf{p}^N; t)$. The measurable observable $A(t)$ is then the mean value of  $\hat A\equiv \hat A(\mathbf{r}^N,\mathbf{p}^N)$ according to Eq. (\ref{meanvalueofdynvar}). 
For example, let the $N$ particles of our system interact only through pairwise forces, whose interaction potential between two particles at positions $\mathbf{r}$ and  $\mathbf{r}'$ is denoted by $u(\mathbf{r}, \mathbf{r}')$, and which are also subjected to an external field such that the potential energy of one particle at position   $\mathbf{r}$ is $\Psi (\mathbf{r})$. Then the total mechanical energy is $\hat U({\bf r}^N,{\bf p}^N)=  \hat K({\bf p}^N) + \hat V({\bf r}^N)$,  with 
\begin{equation}
\hat K({\bf p}^N)= \displaystyle{\sum _{1\le i\le N}} \mathbf{p}_i^2/2M 
\label{kineticenergy}
\end{equation}
being the kinetic energy  and with 
\begin{equation}
\hat V({\bf r}^N)\equiv \sum _{1\le i< j \le N} u(\mathbf{r}_i,\mathbf{r}_j)  + \sum _{1\le i\le N} \Psi(\mathbf{r}_i), 
\label{energyeq0}
\end{equation}
being the potential energy. Then, the total internal energy $U(t)$ of the system is 
\begin{equation}
U(t)= \langle \hat K({\bf p}^N) + \hat V({\bf r}^N) \rangle =  U^{id}(t) + U^{ex}(t).
\label{totalinternalenergy}
\end{equation}

In this equation the ideal term is defined as $U^{id}(t) \equiv \int \hat K({\bf p}^N) \ \mathcal{P}_N({\bf r}^N, {\bf p}^N; t)\ d{\bf r}^N d{\bf p}^N$, which can be rewritten as 
\begin{eqnarray}
U^{id}(t) &\equiv&   \int  \left[\displaystyle{\sum _{1\le i\le N}}\mathbf{p}_i^2/2M\right]\   \mathcal{P}_N({\bf r}^N, {\bf p}^N; t)\ d{\bf r}^N  d{\bf p}^N  \nonumber  \\ 
&=& \displaystyle{\sum _{1\le i\le N}} \int d{\bf r}_i \int d{\bf p}_i \  [\mathbf{p}_i^2/2M]  \left[ \int \mathcal{P}_N({\bf r}_i, {\bf r}^{N-1}, {\bf p}_i ,{\bf p}^{N-1}; t)\ d{\bf r}^{N-1}\ d{\bf p}^{N-1}  \right]      \nonumber  \\ 
&\equiv& \displaystyle{\sum _{1\le i\le N}} \int d{\bf r}_i \int d{\bf p}_i \  [\mathbf{p}_i^2/2M]  \ \mathcal{P}_1({\bf r}_i,  {\bf p}_i ; t)      \nonumber  \\ 
\label{energyeq01}
\end{eqnarray} 
Here $\mathcal{P}_1({\bf r},  {\bf p} ; t) \equiv  \int \mathcal{P}_N({\bf r},{\bf r}_2,\ ..., {\bf r}_N; {\bf p},{\bf p}_2,\ ..., {\bf p}_N; t)\ d{\bf r}_2\ ... d{\bf r}_N d{\bf p}_2\ ... d {\bf p}_N$ is the reduced one-particle PDF, describing the probability that one of the $N$ particles is at position {\bf r} with momentum {\bf p} at time $t$. Eq. (\ref{energyeq01}) can be further rewritten as 
\begin{eqnarray}
U^{id}(t) &\equiv&    \displaystyle{\sum _{1\le i\le N}} \int d{\bf r}_i \int d{\bf p}_i \  [\mathbf{p}_i^2/2M]   \mathcal{P}_1({\bf r}_i,  {\bf p}_i ; t) \int d{\bf r} \ \delta({\bf r}-{\bf r}_i)     \nonumber  \\   
&=& \int d{\bf r} \ \left\{ \displaystyle{\sum _{1\le i\le N}}  \int d{\bf r}_i \int d{\bf p}_i \  [\mathbf{p}_i^2/2M] \delta({\bf r}-{\bf r}_i)\   \mathcal{P}_1({\bf r},  {\bf p}_i ; t)  \right\}         \nonumber  \\   
&=& \int d{\bf r} \ \overline{k}({\bf r},t).          \nonumber  \\   
\label{energyeq2}
\end{eqnarray}

The function $\overline{k}({\bf r},t)$, defined  above as 
\begin{equation}
\overline{k}({\bf r},t)\equiv \left\{ \displaystyle{\sum _{1\le i\le N}}  \int d{\bf r}_i \int d{\bf p}_i \  [\mathbf{p}_i^2/2M] \delta({\bf r}-{\bf r}_i)\   \mathcal{P}_1({\bf r},  {\bf p}_i ; t)  \right\}, 
\label{kderyt}
\end{equation} 
can be identified with the local mean kinetic energy density (per unit volume). Thus, denoting by $\overline{n}({\bf r},t)$ the local mean particle number density (per unit volume), then the ratio $\overline{k}({\bf r},t)/\overline{n}({\bf r},t)$ is the local mean kinetic energy density \emph{per particle}. This, however, is essentially the molecular definition of the \emph{local temperature} $T({\bf r} ; t)$. More precisely, $T({\bf r} ; t)$ will be defined as 
\begin{equation}
 T({\bf r} ; t)= \left(\frac{2}{3k_B}\right) \frac{\overline{k}({\bf r},t)}{\overline{n}({\bf r},t)},
\label{definitionoftemperature}
\end{equation}  
so that in general we can write $U^{id}(t)$ as
\begin{equation}
U^{id}(t)= \frac{3}{2}k_B\int d{\bf r} \ \overline{n}({\bf r},t) T({\bf r},t). 
\label{idealinternalenergygeneral}
\end{equation}  
For future reference, we shall denote by $\beta({\bf r},t)$ the inverse  local mean kinetic energy, 
\begin{equation}
\beta({\bf r},t)=1/k_BT({\bf r},t).
\label{defbeta0}
\end{equation}

Under thermodynamic equilibrium conditions, where $T({\bf r},t)=T$ is constant and uniform, Eq. (\ref{idealinternalenergygeneral}) becomes the ordinary ideal thermal equation of state, 
\begin{equation}
U^{id}_{eq}= \frac{3}{2}Nk_BT.
\label{idealinternalenergyequilibrium}
\end{equation}  
Let us point out, however, that we may also consider other idealized but non-equilibrium conditions, such as assuming  $T({\bf r},t)$ to be uniform but not constant, $T({\bf r},t)=T(t)$, with the time-dependent temperature $T(t)$ controlled by means of thermal reservoirs (assuming, of course, infinite thermal conductivity), thus becoming a control parameter. Under these conditions, we can express $U^{id}(t)$ as 
\begin{equation}
U^{id}(t)= (3/2)Nk_BT(t),
\label{idealinternalenergy}
\end{equation}  
where we define the time-dependent molecular temperature $T(t)$ as 
\begin{equation}
k_BT(t)\equiv   \langle \mathbf{p}^2(t)/3M\rangle,
\label{molectemp}
\end{equation}  
where $\mathbf{p}(t)$ is the momentum of one representative particle.

The last term of Eq. (\ref{totalinternalenergy}) is the \emph{structural} (or \emph{``excess''}) contribution to the internal energy $U$ of the system, $U^{ex}(t)\equiv  \langle \hat V({\bf r}^N (t))\rangle  = \int \hat V({\bf r}^N) \ \mathcal{P}_N({\bf r}^N, {\bf p}^N; t)\ d{\bf r}^N\ d{\bf p}^N  = \int \hat V({\bf r}^N) \ \mathcal{P}_N({\bf r}^N; t)\ d{\bf r}^N$ (with $\mathcal{P}_N({\bf r}^N; t)$ being the reduced PDF defined above), 
so that 
\begin{eqnarray}
U^{ex}(t) = \langle \hat V (t) \rangle  &=&  \int   \left[ \sum _{1\le i< j \le N} u(\mathbf{r}_i,\mathbf{r}_j) + \sum _{1\le i\le N} \Psi(\mathbf{r}_i) \right]     \ \mathcal{P}_N({\bf r}^N; t)\ d{\bf r}^N  \nonumber  \\ 
& =&  \sum _{1\le i< j \le N}  \int   u(\mathbf{r}_i,\mathbf{r}_j)   \ \mathcal{P}_N({\bf r}^N; t)\ d{\bf r}^N\nonumber  \\ 
& &  + \sum _{1\le i \le N}  \int   \Psi(\mathbf{r}_i)  \ \mathcal{P}_N({\bf r}^N; t)\ d{\bf r}^N.   
\label{energyeq1}
\end{eqnarray}
Since each of the $N(N-1)/2$ terms of the first sum contribute equally, and each of the $N$ terms of the second sum also  contribute equally,  this expression can also be written as 
\begin{equation}
U^{ex}(t)  = \frac{1}{2}\int   u(\mathbf{r}_1,\mathbf{r}_2)  n^{(2)}(\mathbf{r}_1, \mathbf{r}_2;t)   d \mathbf{r}_1 d \mathbf{r}_2 + \int   \Psi (\mathbf{r}_1)  n^{(1)}(\mathbf{r}_1;t)   d \mathbf{r}_1, 
\label{energyeq2}
\end{equation}

\medskip
\noindent where $n^{(1)}(\mathbf{r}_1;t)$ and $n^{(2)}(\mathbf{r}_1, \mathbf{r}_2;t)$ are the one-particle and the two-particles time-dependent density distribution functions. In general, the $\nu$-particle \emph{density distribution functions} ($\nu$-DDF) are defined (for $1\le \nu \le N$) as 
\begin{equation}
 n^{(\nu)}(\mathbf{r}_1, \mathbf{r}_2, ..., \mathbf{r}_\nu;t) \equiv \frac{N!}{(N-\nu)!} \int  \ \mathcal{P}_N({\bf r}^N; t)   \ d \mathbf{r}_{\nu+1}\   \ d \mathbf{r}_{\nu+2}\ ....\ d \mathbf{r}_{N}, 
\label{nparticledensity}
\end{equation}
normalized such that $\int  n^{(\nu)}(\mathbf{r}_1, \mathbf{r}_2, ..., \mathbf{r}_\nu;t) d\mathbf{r}_1 d\mathbf{r}_2, ..., d\mathbf{r}_\nu=N!/(N-\nu)! $. Using now Eqs. (\ref{idealinternalenergygeneral})  and  (\ref{energyeq2})  in Eq. (\ref{totalinternalenergy}), we finally get 
\begin{equation}
U(t)  = \frac{3}{2}k_B\int d{\bf r} \ n^{(1)}({\bf r},t) T({\bf r},t)+\frac{1}{2}\int   u(\mathbf{r}_1,\mathbf{r}_2)  n^{(2)}(\mathbf{r}_1, \mathbf{r}_2;t)   d \mathbf{r}_1 d \mathbf{r}_2 + \int   \Psi (\mathbf{r}_1)  n^{(1)}(\mathbf{r}_1;t)   d \mathbf{r}_1. 
\label{energyeqgeneral}
\end{equation}

All of the equations above are either general definitions or exact relationships among the defined objects which, remarkably, do not invoke any particular condition on the (generally non-equilibrium and time-dependent) PDF $\mathcal{P}_N({\bf q^N}, \mathbf{p}^N; t)$ or on its reduced versions $\mathcal{P}_N({\bf p}^N; t)$ and $\mathcal{P}_N({\bf r}^N; t)$. For the assumed potential energy in Eq. (\ref{energyeq0}), for example, the expression in Eq. (\ref{energyeqgeneral}) is the most general and exact form of the so-called \emph{energy equation},  which expresses the macroscopic thermodynamic property $U(t) = \langle \hat K (t) +\hat V (t)\rangle $, in terms of the one- and two-particles density distribution functions $n^{(1)}(\mathbf{r}_1;t)$ and $n^{(2)}(\mathbf{r}_1, \mathbf{r}_2;t)$.

In fact, one can consider additional specific circumstances, and adapt Eq. (\ref{energyeq2}) accordingly. For example,  in the absence of external fields, $ \Psi (\mathbf{r})= 0$, and for fluids with radially-symmetric interactions, i.e., if $u(\mathbf{r},\mathbf{r}')=u(\mid\mathbf{r}-\mathbf{r}'\mid)$, the symmetry condition of spatial uniformity and isotropy imply that $n^{(1)}(\mathbf{r};t) = n\equiv N/V$ and $n^{(2)}(\mathbf{r}, \mathbf{r}';t) = n^{(2)}(\mid\mathbf{r}-\mathbf{r}'\mid;t) =n\delta (\mathbf{r} - \mathbf{r}') + n^2 g(\mid\mathbf{r}-\mathbf{r}'\mid;t)$, where the function $g(r;t)$ is the time-dependent non-equilibrium \emph{radial distribution function}, defined in an entirely analogous fashion as its equilibrium counterpart (see, for instance, Eqs.(2.5.8) and (2.5.9) of Ref. \cite{hansen}). Under these specific conditions, the energy equation (\ref{energyeq2}) may be rewritten in its most familiar form, 
\begin{equation}
\frac{U^{ex}(t)}{N}  = \frac{n}{2}\int   u(r)  g(r;t)   d^3 \mathbf{r}.
\label{energyeq3}
\end{equation} 

This expression for $U^{ex}(t)$, and its more general version in Eq. (\ref{energyeq2}), are valid at non-equilibrium and equilibrium conditions, since their derivation never  assumed thermodynamic equilibrium, i.e., never employed Eq. (\ref{canonicalpdf}). The challenge now is how to determine the non-equilibrium structural properties $n^{(1)}(\mathbf{r}_1;t)$ and $n^{(2)}(\mathbf{r}_1, \mathbf{r}_2;t)$. In the following discussion  we describe how this challenge has been addressed by the NE-SCGLE theory.

\subsection{The Ornstein-Zernike equation, a thermodynamic equilibrium condition.} \label{subsection2.3}   

One of the most useful concepts in the equilibrium theory of liquids is, indeed, the Ornstein-Zernike (OZ) equation. Let us consider the equilibrium mean value $n^{eq}(\mathbf{r}) \equiv \langle \hat n(\mathbf{r}) \rangle^{eq}$  of the local number density $\hat n(\mathbf{r}) \equiv \sum_i^N \delta (\mathbf{r}-\mathbf{r}_i)$ and the corresponding covariance $\sigma^{eq}(\textbf{r},\textbf{r}')\equiv \langle [\hat n(\mathbf{r})-n^{eq}(\mathbf{r})] [\hat n(\mathbf{r}')-n^{eq}(\mathbf{r}')]\rangle^{eq}=n^{(2)}_{eq}(\mathbf{r}, \mathbf{r}') - n^{eq}(\mathbf{r})n^{eq}(\mathbf{r}')$. We notice that we can split $n^{(2)}_{eq}(\mathbf{r}, \mathbf{r}')$ into its \emph{self} and \emph{distinct} parts, $n^{(2)}_{eq}(\mathbf{r}, \mathbf{r}') = n^{eq}(\mathbf{r})\delta ({\bf r}-\textbf{r}')+n^{eq}(\mathbf{r})n^{eq}(\mathbf{r}')g ({\bf r},\textbf{r}')$, where $g ({\bf r},\textbf{r}')$ is the pair distribution function. This allows us to write the covariance as 
\begin{equation} 
\sigma^{eq}(\textbf{r},\textbf{r}')  = n^{eq}(\mathbf{r})\delta ({\bf r}-\textbf{r}')+n^{eq}(\mathbf{r})n^{eq}(\mathbf{r}')h ({\bf r},\textbf{r}'),
\label{sigmaeqselfplusdistinct}
\end{equation}
where $h ({\bf r},\textbf{r}')\equiv g ({\bf r},\textbf{r}')-1$ is the  {\it total correlation function}. 

Just like the mean value $n^{eq}(\mathbf{r})$ is determined by the well known chemical equilibrium condition $\nabla \mu [{\bf r};n]=0$ (where the functional $\mu [{\bf r};n]$ is the chemical potential), the covariance  $\sigma^{eq}(\textbf{r},\textbf{r}')$ is determined by its corresponding thermodynamic equilibrium condition,
\begin{equation} 
\int \sigma^{eq}(\textbf{r},\textbf{r}') \mathcal{E}[{\bf r}',\textbf{r}'' ] d {\bf r}' = \delta ({\bf r}-\textbf{r}''),
\label{sigmaxeeqi}
\end{equation}
where the stability function $\mathcal{E}[{\bf r}',\textbf{r}' ]$ is defined as the functional derivative
\begin{equation} 
\mathcal{E}[{\bf r},\textbf{r}']\equiv \left(\frac {\delta \beta\mu [{\bf r};n]}{\delta n(\textbf{r}')}\right)_{n=n^{eq}}.
\label{stabilityfunctione1}
\end{equation}
Under conditions of spatial uniformity and isotropy, $\sigma^{eq}(\textbf{r},\textbf{r}') =\sigma^{eq}(\mid {\bf r}-{\bf r'}\mid)$ and $\mathcal{E}[{\bf r},\textbf{r}']= \mathcal{E}(\mid {\bf r}-{\bf r'}\mid)$, and the thermodynamic equilibrium condition for $\sigma^{eq}(\textbf{r},\textbf{r}')$ reads, in terms of the Fourier transforms $\sigma^{eq}(k)$ and $\mathcal{E}(k)$, of $\sigma^{eq}(r)$ and $\mathcal{E}(r)$, respectively, as 
\begin{equation} 
\sigma^{eq}(k) \mathcal{E}(k) = 1.
\label{sigmaxeeqift}
\end{equation}

In general, however, since the chemical potential is the sum of its ideal plus its ``excess'' contributions, $\mu [{\bf r};n] = k_BT \ln n(\mathbf{r}) + \mu^{ex} [{\bf r};n]$,  the function $\mathcal{E}[{\bf r}',\textbf{r}' ]$ can also be written as \cite{hansen}  
\begin{equation} 
\mathcal{E}[{\bf r},\textbf{r}'] = \frac{ \delta ({\bf r}-\textbf{r}')}{n^{eq}(\mathbf{r})} + \left(\frac {\delta \beta\mu^{ex} [{\bf r};n]}{\delta n(\textbf{r}')}\right)_{n=n^{eq}}  \equiv \frac{ \delta ({\bf r}-\textbf{r}')}{n^{eq}(\mathbf{r})} - c(\mathbf{r}, \mathbf{r}'), \label{stabilityfunctione2}
\end{equation}
with $c(\mathbf{r}, \mathbf{r}') \equiv -\left(\delta \beta\mu^{ex} [{\bf r};n]/\delta n(\textbf{r}')\right)_{n=n^{eq}}$ being the direct correlation function. It is then not difficult to see, using Eqs. (\ref{sigmaeqselfplusdistinct}) and (\ref{stabilityfunctione2}) in the thermodynamic equilibrium condition in Eq. (\ref{sigmaxeeqi}), that the latter equation is nothing but the well-known Ornstein-Zernike equation, 
\begin{equation} 
h({\bf r},{\bf r'})=c({\bf r},{\bf r'})+ \int d^3r''c({\bf r},{\bf r''})n^{eq}({\bf r''})h({\bf r''},{\bf r'}). 
\label{ozeq0} 
\end{equation}

Thus,  the question of what is the non-equilibrium extension of the Ornstein-Zernike equation, can now be answered: there is no such non-equilibrium extension, since the Ornstein-Zernike equation itself is an essential signature of thermodynamic equilibrium \cite{brader1}. In fact, written as the equilibrium condition in Eq. (\ref{sigmaxeeqi}), it determines the covariance $\sigma^{eq}(\textbf{r},\textbf{r}')$ in terms of the thermodynamic stability function $\mathcal{E}[{\bf r},\textbf{r}']$. Hence, an alternative (and more relevant) question that makes sense is: which are the equations that determine the \emph{non-equilibrium} mean particle number density $\overline{n}(\textbf{r},t)$ and the covariance $\sigma(\textbf{r},\textbf{r}';t)$ (or $\sigma(k;\textbf{r},t)$)?. Fortunately, we are presently in the position to also answer this question: the equations whose solution determines the  non-equilibrium properties $\overline{n}(\textbf{r},t)$ and $\sigma(k;\textbf{r},t)$ are precisely the central equations of the NE-SCGLE theory, namely,  Eqs. (\ref{difeqdlp}) and (\ref{relsigmadif2p}) below.

\subsection{Non-equilibrium statistical mechanics  of liquids: the NE-SCGLE theory.} \label{subsection2.4}

Focusing on thermodynamic  equilibrium was a perfectly reasonable starting point since, in principle, one might theoretically calculate  $n^{(1)}_{eq}(\mathbf{r}_1)$ and $n^{(2)}_{eq}(\mathbf{r}_1, \mathbf{r}_2)$  from the fundamental expression for $\mathcal{P}_N^{eq}(\mathbf{r}^N,\mathbf{p}^N)$ in Eq. (\ref{canonicalpdf}). This, in fact, led to the development of the integral equations formalism and the density functional theory \cite{evans,mcquarrie,hansen} in the early stages of the statistical mechanical description of liquids. Nowadays, however, the growing interest throughout the  natural sciences for the understanding of matter out-of-equilibrium, demands revising the original limitations of these approaches. Thus, it  makes perfect sense to revisit the foundations of the statistical mechanical theory of liquids, to launch a similar effort to establish the fundamental general principles and the specific and practical theoretical approaches to predict and calculate not only the non-equilibrium structural properties $n^{(1)}(\mathbf{r}_1;t)$ and $n^{(2)}(\mathbf{r}_1, \mathbf{r}_2;t)$, but all the other measurable physical properties that characterize the behavior of non-equilibrium liquids.

A major pioneering contribution of the Mexican statistical physics community to this endeavor has been the development of the \emph{non-equilibrium self-consistent generalized Langevin equation} (NE-SCGLE) theory of irreversible processes in liquids \cite{nescgle1,nescgle2,nescgle3}, whose central elements are, precisely, the general time-evolution equations of the non-equilibrium structural properties $n^{(1)}(\mathbf{r}_1;t)$ and $n^{(2)}(\mathbf{r}_1, \mathbf{r}_2;t)$. These are Eqs. (4.1) and (4.2) of Ref. \cite{nescgle1}, which we rewrite here for easy reference in terms of the mean particle number density $\overline{n}(\textbf{r},t)\equiv n^{(1)}(\mathbf{r};t)$ and the covariance 
$\sigma(\textbf{r},\textbf{r}';t)\equiv n^{(2)}(\mathbf{r}, \mathbf{r}';t) - n^{(1)}(\mathbf{r};t)n^{(1)}(\mathbf{r}';t)$ (for specific details, the reader is referred to Refs. \cite{nescgle1,nonlinonsmach}). The first of these equations reads 
\begin{equation} 
\frac{\partial \overline{n}(\textbf{r},t)}{\partial
t} = D^0{\nabla} \cdot b(\textbf{r},t)\overline{n}(\textbf{r},t)
\nabla \hat{\beta}\mu[{\bf r};\overline{n}(t),\overline{T}(t)], 
\label{difeqdlp}
\end{equation}
whereas the second is written in terms of the Fourier transform (FT) $\sigma(k;\textbf{r},t)$ of the globally non-uniform but locally (approximately) homogeneous covariance $ \sigma(\mid \textbf{x}\mid;\textbf{r},t) \equiv \sigma (\textbf{r},\textbf{r} + \textbf{x};t)$, as
\begin{eqnarray}
\frac{\partial \sigma(k;\textbf{r},t)}{\partial t} =  -2k^2 D^0 \overline{n}(\textbf{r},t) b(\textbf{r},t)
\mathcal{E}(k;\overline{n}(\textbf{r},t)) \left[\sigma(k;\textbf{r},t)- \frac{1}{\mathcal{E}(k;\overline{n}(\textbf{r},t))}\right].
\label{relsigmadif2p}
\end{eqnarray}
As explained in Section 5 of Ref. \cite{nonlinonsmach}, in these equations $D^0$ is the particles' short-time self-diffusion coefficient \cite{pusey} and $b(\textbf{r},t)$ is their local reduced mobility. In addition, $\mu[{\bf r};\overline{n}(t),\overline{T}(t)]$ is the chemical potential per particle,  defined as  $\mu[{\bf r};n;T] \equiv  (\delta F[n;T]/\delta n(\textbf{r}))$ evaluated at $n=\overline{n}(t)$, where $F[n;T]$ is the Helmholtz free energy  density \emph{functional}. The second functional derivative of $F[n;T]$ determines the stability matrix $\mathcal{E}[{\bf r},\textbf{r}';n;T]\equiv \beta (\delta^2 F[n;T]/\delta n(\textbf{r})\delta n(\textbf{r}')) = \left( {\delta \beta\mu [{\bf r};\overline{n}(t);T]}/{\delta n(\textbf{r}')}\right)$. The function $\mathcal{E}(k;\overline{n}(\textbf{r},t))$ that appears in Eq. (\ref{relsigmadif2p}) is the FT of $\mathcal{E}[{\bf r},\textbf{r} + \textbf{x};n;T]\equiv \left[ {\delta \beta\mu [{\bf r};n;T]}/{\delta n(\textbf{r} + \textbf{x})}\right]$.


The NE-SCGLE theory originates from a generalization of Onsager's description of irreversible processes  \cite{onsager1,onsager2} and fluctuations \cite{onsagermachlup1,onsagermachlup2}, to genuine non-equilibrium and non-linear conditions \cite{nescgle1}. Applied to the description of irreversible processes in liquids, this canonical and abstract formalism becomes a generic first-principles theory of its structure and dynamics, at equilibrium and during the non-stationary processes of equilibration or aging. Here, however, we do not mean to review these theoretical advances, but refer the reader to a related separate work \cite{nonlinonsmach}, which summarizes the  fundamental basis and origin of the NE-SCGLE theory (and hence of these equations), and also guides the reader through the pertinent references. Instead, at this point we would like to use these equations as the starting point of other relevant discussions. In particular, let us now discuss the significance of the Ornstein-Zernike equation in the discussion of non-equilibrium phenomena.


\subsection{Equilibrium vs. arrested states.} \label{subsection2.5}   

Even before solving Eqs. (\ref{difeqdlp}) and (\ref{relsigmadif2p}), these equations reveal an important general feature. To see this, let us first discuss their stationary limit. As already discussed above, thermodynamic equilibrium states correspond to the stationary solutions $\overline{n}^{eq}(\textbf{r})$ and $\sigma^{eq} (k;\textbf{r})$, that satisfy the equilibrium conditions $\nabla \beta\mu[{\bf r};\overline{n}^{eq}]=0$ and $\sigma^{eq} (k;\textbf{r})\mathcal{E}(k;\overline{n}^{eq}(\textbf{r})) =1$ (see Eq. (\ref{sigmaxeeqift})), and which guarantee stationarity. Eqs.  (\ref{difeqdlp}) and (\ref{relsigmadif2p}), however, also predict the possibibility of another set of stationary solutions, whose stationarity is guaranteed because they fulfill the dynamic arrest condition, $\displaystyle{\lim_{t\to \infty} b(\textbf{r},t)=0}$, without satisfying the equilibrium conditions, i.e., without maximizing entropy. This second set of solutions describes non-equilibrium stationary states of matter, corresponding to glasses, gels, and other non-equilibrium amorphous solids.

The existence of this universal catalog of dynamically arrested states of matter was first envisaged not through the analysis of thermodynamic or structural properties (such as $\overline{n}^{eq}(\textbf{r})$ and $\sigma^{eq} (k;\textbf{r})$), but through a dynamical criterion based on the mode-coupling theoretical analysis of the asymptotic long-$\tau$ limit of the Fourier transform $C^{eq}(k,\tau)$ of the \emph{equilibrium} time-dependent correlation function $C^{eq}(\mid\mathbf{r}-\mathbf{r}'\mid,\tau)\equiv \langle [\hat n(\mathbf{r},\tau)-n^{eq}(\mathbf{r})] [\hat n(\mathbf{r}';0)-n^{eq}(\mathbf{r}')]\rangle^{eq}$ (see, for example, Section 3.2 of Ref. \cite{goetze4}, or Section 4.3 of G\"otze's book \cite{goetzebook}). If $C^{eq}(k,\tau)$ decays with $\tau$ to zero, the system will reach a thermodynamic equilibrium state, whereas if  $C^{eq}(k,\tau)$ decays to a finite value, the system will become dynamically arrested. This criterion allows us to partition the state space of a given system (for example, the temperature-density plane) into two mutually-excluding regions, the liquid (or ``ergodic'') and the glass (or ``non-ergodic'') regions. The resulting diagram is  referred to as \emph{glass transition} \cite{sperl1} (or \emph{dynamic arrest} \cite{pedroatractivos}) diagram. Unfortunately, being based on an equilibrium theory of dynamic properties \cite{elizondo_mct},  this glimpse of a non-equilibrium scenario is limited in several respects, most notoriously by its inability to describe time-dependent relaxation processes.

To escape from this limitation, let us now discuss the full time-dependent solution of Eqs.  (\ref{difeqdlp}) and (\ref{relsigmadif2p}), with arbitrary initial conditions  $\overline{n}^{0}(\textbf{r})$ and $\sigma^{0} (k;\textbf{r})$. This $t$-dependent solution describes the full relaxation of a liquid prepared at that initial state, but instantaneously quenched at $t=0$ to an arbitrary final temperature and density. The temporal evolution of $\overline{n}(\textbf{r},t)$ and $\sigma(k;\textbf{r},t)$ narrates a full story, from its known beginning ($t=0$) to its unknown end ($t\to \infty$). A relevant question is, then, what will be the end of this story?, i.e., for arbitrarily-given  $\overline{n}^{0}(\textbf{r})$ and $\sigma^{0} (k;\textbf{r})$, what will be the value of $\overline{n}(\textbf{r},t\to \infty)$ and $\sigma(k;\textbf{r},t\to \infty)$?. From a purely equilibrium statistical thermodynamic perspective, our guess will be that the only possible end of this story will be to reach thermodynamic equilibrium, i.e., that $\overline{n}(\textbf{r},t\to \infty)=\overline{n}^{eq}(\textbf{r})$ and $\sigma(k;\textbf{r},t\to \infty)=\sigma^{eq} (k;\textbf{r})$, since the central dogma of equilibrium statistical mechanics is that any stationary state of matter must satisfy the maximum entropy principle, i.e., must belong to the universal catalog of equilibrium states. 

The kinetic perspective of the NE-SCGLE theory challenges this dogma, and provides the route of escape from  the corresponding limiting constraint. This starts with Eqs.  (\ref{difeqdlp}) and (\ref{relsigmadif2p}), which announce the alternative possibility that stationary solutions $\overline{n}(\textbf{r},t\to \infty)=\overline{n}^{a}(\textbf{r})$ and $\sigma(k;\textbf{r},t\to \infty)=\sigma^{a} (k;\textbf{r})$  exist, which do not maximize the entropy, but satisfy instead a kinetic condition of dynamic arrest, in the present case the vanishing of the molecular mobility, $\displaystyle{\lim_{t\to \infty} b(\textbf{r},t)=0}$. These ``new'' stationary solutions of Eqs.  (\ref{difeqdlp}) and (\ref{relsigmadif2p}) constitute the second universal category of states of matter, whose existence was also announced by MCT dynamic arguments for equilibrium liquids, but which could never had been conceived or discovered from a purely equilibrium statistical thermodynamic perspective. In this sense, this is an unprecedented and remarkable revelation of the NE-SCGLE theory, that originates from the non-linearity of  Eqs.  (\ref{difeqdlp}) and (\ref{relsigmadif2p}), which innocently hide the fact that the transport coefficients are in reality state functions; in the present case,  that $b(\textbf{r},t)$ is in reality a \emph{functional} of $\overline{n}(\textbf{r},t)$ and $\sigma(k;\textbf{r},t)$ \cite{peredo1}.

\subsection{The ``kinetic equation of state''.} \label{subsection2.6}   

Of course, in order to prove the rather strong statements above, we need the ``kinetic equation of state'', i.e., the functional dependence of $b(\textbf{r},t)$ on $\overline{n}(\textbf{r},t)$ and $\sigma(k;\textbf{r},t)$, and to actually solve Eqs.  (\ref{difeqdlp}) and (\ref{relsigmadif2p}) to exhibit the arrested solutions $\overline{n}^{a}(\textbf{r})$ and $\sigma^{a} (k;\textbf{r})$ in concrete specific examples. This program has been carried out in many convincing examples, although within an important simplifying approximation, in which one imposes the condition of spatial homogeneity, so that  $\overline{n}(\textbf{r};t)\approx n=N/V$ and $\sigma(k;\textbf{r},t)\approx \sigma (k;t)= n S(k;t)$, where $S(k;t)$ is the non-equilibrium structure factor. This simplifies Eqs.  (\ref{difeqdlp}) and (\ref{relsigmadif2p}) to become a single time-evolution equation for $S(k;t)$, namely, 
\begin{equation}
\frac{\partial S(k;t)}{\partial t} = -2k^2 D^0
b(t)n\mathcal{E}(k;n,T) \left[S(k;t)
-1/n\mathcal{E}(k;n,T)\right]. \label{dsktenst}
\end{equation}

Thus, for ``kinetic equation of state'' we now mean the functional dependence of the  time-dependent mobility function $b(t)$ on the time-dependent structure factor $S(k;t)$ \cite{peredo1}. The equations determining this functional dependence start with Einstein's relation, 
\begin{equation}
b(t)= [1+\int_0^{\infty} d\tau\Delta{\zeta}^*(\tau; t)]^{-1},
\label{bdtp}
\end{equation}
between the mobility $b(t)$ and the $t$-evolving, $\tau$-dependent friction function $\Delta{\zeta}^*(\tau; t)$, for which it is possible to derive the following approximate expression \cite{nescgle1},
\begin{equation}
\begin{split}
  \Delta \zeta^* (\tau; t)= \frac{D_0}{24 \pi ^{3}n}  \int d {\bf k}\ k^2 \left[\frac{ S(k;t)-1}{S(k; t)}\right]^2  \\ \times F(k,\tau; t)F_S(k,\tau; t),
\end{split}
\label{dzdtquench}
\end{equation}
in terms of $S(k; t)$, and of the non-equilibrium intermediate scattering function (NEISF) $F(k,\tau; t)\equiv N^{-1}  \langle \delta n(\mathbf{k},t+\tau) \delta n(-\mathbf{k},t)\rangle$, where  $\delta n(\mathbf{k},t)$ is the FT of the thermal fluctuations $\delta n(\mathbf{r},t)\equiv \hat{n}(\mathbf{r},t)-n$ of the local number density  $n(\mathbf{r},t)$ at time $t$. The self-NEISF $F_S(k,\tau; t)$, in turn, is defined as $F_S(k,\tau; t)\equiv  \langle \exp \left[    i\mathbf{k}\cdot \Delta \mathbf{r}_T(t,\tau)\right]\rangle$, with $\Delta \mathbf{r}_T(t,\tau) \equiv  \left[ \mathbf{r}_T(t+\tau)-\mathbf{r}_T(t)\right]$ being the displacement of one particle considered as a  tracer. 

The previous equations are complemented by the corresponding memory-function equations for $F(k,\tau; t)$ and $F_S(k,\tau; t)$, written approximately, in terms of their Laplace transforms (LT) $F(k,z; t)$ and $F_S(k,z; t)$, as
\begin{gather}\label{fluctquench}
 F(k,z; t) = \frac{S(k; t)}{z+\displaystyle{\frac{k^2D^0 S^{-1}(k;
t)}{1+\lambda (k)\ \Delta \zeta^*(z; t)}}},
\end{gather}
and
\begin{gather}\label{fluctsquench}
 F_S(k,z; t) = \frac{1}{z+\displaystyle{\frac{k^2D^0 }{1+\lambda (k)\ \Delta \zeta^*(z; t)}}},
\end{gather}
where the memory functions of both, $F(k,z; t)$ and $F_S(k,z; t)$, were approximated by the product $\lambda (k)\ \Delta \zeta^*(z; t)$. In these equations $\Delta \zeta^*(z; t)$ is the LT of $\Delta \zeta^*(\tau; t)$ and  $\lambda (k)\equiv 1/[1+( k/k_{c}) ^{2}]$ is an ``interpolating function" \cite{todos1}, with $k_{c}$ being an empirically determined cutoff wave-vector. For the hard sphere liquid, for example,  $k_{c}= 1.305(2\pi)/\sigma$, with $\sigma$ being the hard-core particle diameter. 

Since the function $\mathcal{E}(k;n,T)$ is considered known, Eqs. (\ref{dsktenst})-(\ref{fluctsquench}) constitute a closed system of equations whose solution determines $S(k;t)$,   $b(t)$, $\Delta{\zeta}^*(\tau; t)$, $F(k,\tau; t)$, and $F_S(k,\tau; t)$. These equations thus embody the kinetic equation of state we referred to above. They, in addition, constitute the mathematical summary of the NE-SCGLE theory in its simplest practicable version. Along the short history of this theory (a bit more than one decade), these equation have been solved for a number of physically significant model systems. Before the NE-SCGLE theory,  the results of the application of mode-coupling theory, and of the equilibrium SCGLE theory (summarized by Eqs. (\ref{dzdtquench})-(\ref{fluctsquench}) with $S(k; t)$ replaced by $S^{eq}(k)$ \cite{scgle1,scgle2,todos1}), represented the state of the art in the first-principles theoretical description of dynamic arrest and amorphous solidification \cite{elizondo_mct}. The scenario predicted by these equilibrium theories excluded any reference to the $t$-dependent non-equilibrium transients (such as the aging of quenched glass-forming liquids), whose description given by the solution of the full  NE-SCGLE equations constitute the new state of the art. For example, the NE-SCGLE theory allows the first-principles prediction of the time-dependent non-equilibrium phase diagram of simple glass- and gel-forming liquids \cite{zepeda}, the theoretical counterpart of the so-called time-temperature-transformation (TTT) diagrams \cite{atlasttd,nakashima}, which are the empirical non-equilibrium extension of the ordinary equilibrium phase diagrams.

\subsection{Partial summary.} \label{subsection2.7}   

In this section we have discussed some aspects of a theoretical methodology  employed to extend to non-equilibrium, well-known concepts of the equilibrium theory of liquids. Such methodology consists of revising the derivation of a given theoretical result, to see if the restriction to equilibrium was really necessary. This was illustrated with the energy equation, Eq. (\ref{energyeq2}), whose validity at non-equilibrium conditions was easily demonstrated. The same illustrated methodology, however, will find a more relevant application in Section \ref{section5} of this work, which describes the derivation of an expression for the non-equilibrium dynamic shear viscosity $\eta(\omega;t)$ of a colloidal liquid in terms of the non-equilibrium structure factor $S(k;t)$ and ISFs $F(k,\tau; t)$ and $F_S(k,\tau; t)$. In both cases, one expresses one macroscopic property ($U^{ex}(t)$ or $\eta(\omega;t)$) in terms of structural ($n^{(1)}(\mathbf{r}_1;t)$, $n^{(2)}(\mathbf{r}_1, \mathbf{r}_2;t)$, $g(r;t)$, and $S(k;t)$) and/or dynamic ($F(k,\tau; t)$ and $F_S(k,\tau; t)$) properties, whose determination was thus the following challenge.

Hence, in this section we also summarized how such a challenge was addressed by the NE-SCGLE theory, whose central elements are precisely the time-evolution equations of $n^{(1)}(\mathbf{r}_1;t)$ and $n^{(2)}(\mathbf{r}_1, \mathbf{r}_2;t)$ in Eqs. (\ref{difeqdlp}) and (\ref{relsigmadif2p}). From the analysis of these equations we learnt that the equilibrium properties $n^{(1)}_{eq}(\mathbf{r}_1)$ and $n^{(2)}_{eq}(\mathbf{r}_1, \mathbf{r}_2)$ are, indeed, the solution of the well-known thermodynamic equilibrium conditions $\nabla \mu [{\bf r};n]=0$ for $n^{(1)}_{eq}(\mathbf{r}_1)$ and Eq. (\ref{sigmaxeeqi})  for $n^{(2)}_{eq}(\mathbf{r}_1, \mathbf{r}_2)$, which is another manner of writing the Ornstein-Zernike equation. However, in the present section we have also learned that $n^{(1)}_{eq}(\mathbf{r}_1)$ and $n^{(2)}_{eq}(\mathbf{r}_1, \mathbf{r}_2)$ can also be understood as stationary solutions of the the time-evolution equations (\ref{difeqdlp}) and (\ref{relsigmadif2p}) which, crucially, also admit other stationary solutions, denoted as $n^{(1)}_{a}(\mathbf{r}_1)$ and $n^{(2)}_{a}(\mathbf{r}_1, \mathbf{r}_2)$, whose stationarity derive from the kinetic arrest condition $\displaystyle{\lim_{t\to \infty} b(\textbf{r},t)=0}$, and which represent glasses, gels and other non-equilibrium amorphous materials. 

A straightforward manner to prove this statement is to actually exhibit these non-equilibrium arrested solutions. For this, however, we require the kinetic equation of state, i.e., the determination of $b(\mathbf{r};t)$ as a functional of $n^{(1)}(\mathbf{r}_1;t)$ and $n^{(2)}(\mathbf{r}_1, \mathbf{r}_2;t)$.  This is precisely the main contribution of the  NE-SCGLE theory. Within the simplifying constraint that $n^{(1)}(\mathbf{r}_1;t) \approx n$,  the mathematical summary of the NE-SCGLE theory is represented by Eqs. (\ref{dsktenst})-(\ref{fluctsquench}). The solution of  these equations determines  in particular $b(t)$ as a functional of  the non-equilibrium structure factor $S(k;t)$, which is the essence of the kinetic equation of state. 

\section{Non-equilibrium Wertheim-Lovett relation.}\label{section3}

Although approximately, the NE-SCGLE equations (\ref{difeqdlp}) and (\ref{relsigmadif2p}) above, complemented by Eqs. (4.4)-(4.7) of Ref. \cite{nescgle1} (or, within the constraint of spatial uniformity, by Eqs. (\ref{dsktenst})-(\ref{fluctsquench}) of the previous section), address the challenge of determining the non-equilibrium structural properties $n^{(1)}(\mathbf{r}_1;t)$ and $n^{(2)}(\mathbf{r}_1, \mathbf{r}_2;t)$. At equilibrium, however,  $n^{(1)}_{eq}(\mathbf{r}_1)$ and $n^{(2)}_{eq}(\mathbf{r}_1, \mathbf{r}_2)$ are related by a general exact relationship, referred to as the Wertheim-Lovett (WL) relation (see Eq (56) of Ref. \cite{evans}), written as
\begin{equation}
\nabla_1 n^{(1)}_{eq}(\mathbf{r}_1) = -\beta \int d\mathbf{r}_2 \left[n^{(2)}_{eq}(\mathbf{r}_1, \mathbf{r}_2)   + n^{(1)}_{eq}(\mathbf{r}_1)\delta (\mathbf{r}_1- \mathbf{r}_2) - n^{(1)}_{eq}(\mathbf{r}_1)n^{(1)}_{eq}(\mathbf{r}_2) \right] \nabla_2 \Psi (\mathbf{r}_2),
\label{wertheimlovett0}
\end{equation}
where $\Psi (\mathbf{r}_2)$ is the potential of an arbitrary external field. 

In the particular case that $\Psi (\mathbf{r}_2)$  is in reality the pair potential $u(\mathbf{0}, \mathbf{r}_2)$ of the force on one particle at $\mathbf{r}_2$ exerted by another particle fixed at the origin $\mathbf{0}$, this equation reads
\begin{equation}
\nabla_1 n^{(1)}_{eq}(\mathbf{r}_1) = -\beta \int d\mathbf{r}_2\  \sigma^{eq} ({\bf r}_1,{\bf r}_2)  \nabla_2 u(\mathbf{0}, \mathbf{r}_2),
\label{wertheimlovett1}
\end{equation}
where now $n^{(1)}_{eq}(\mathbf{r})$ is the equilibrium mean local density of particles around the particle fixed at the origin (for spherical particles $n^{(1)}_{eq}(\mathbf{r})=ng(r)$, with $g(r)$ being the ordinary radial distribution function), and where $\sigma^{eq} ({\bf r}_1,{\bf r}_2) \equiv \left[n^{(2)}_{eq}(\mathbf{r}_1, \mathbf{r}_2)   + n^{(1)}_{eq}(\mathbf{r}_1)\delta (\mathbf{r}_1- \mathbf{r}_2) - n^{(1)}_{eq}(\mathbf{r}_1)n^{(1)}_{eq}(\mathbf{r}_2) \right]$ is the covariance, also in the presence of the same fixed particle. According to the standard derivation \cite{evans},  the WL relation, as well as other exact relations, such as those involved in the Yvon-Born-Green (YBG) hierarchy \cite{mcquarrie,hansen}, are completely general. Unfortunately, they cannot be taken for granted at non-equilibrium conditions. 

As already discussed, the growing need to describe out-of-equilibrium liquids calls for the identification of the non-equilibrium extension of (some of) these exact results. Thus, in what follows we  describe one derivation of Eq. (\ref{wertheimlovett1}) \cite{faraday}, which does not follow the conventional (equilibrium) route \cite{evans} and indicates the manner to escape from the limitation to equilibrium conditions. In the rest of this section we explain how this leads to the  time-dependent non-equilibrium version of  Eq. (\ref{wertheimlovett1}), namely, 
\begin{equation}
\nabla_1 \overline {n}(\mathbf{r}_1;t) = -\beta \int d\mathbf{r}_2 \ \sigma ({\bf r}_1,{\bf r}_2;t)  \nabla_2 u(\mathbf{0}, \mathbf{r}_2).
\label{wertheimlovett2}
\end{equation}
Such derivation appeals to the generalized Langevin equation (GLE) formalism that results when the  Onsager-Machlup theory of thermal fluctuations is adequately extended to include temporal and spatial non-locality \cite{delrio,faraday}. The elements of this GLE formalism are now summarized.

\subsection{The stationarity theorem and the generalized Langevin equation}

Let us start with a brief discussion of purely mathematical nature, and consider an arbitrary and general \emph{stationary} stochastic process $\textbf{a}(t)$ defined by a stationary ensemble of realizations of the fluctuations $\delta {\textbf a}(t)\equiv {\textbf a }(t)-\langle {\textbf a }\rangle ^{ss}$ around the mean value $\langle {\textbf a }\rangle ^{ss}$ of  $\textbf{a}(t)$. These realizations are generated by the solutions of a linear stochastic equation with additive noise, of the following general form
\begin{equation}
\frac{d \delta {\textbf a }(t)}{dt}=   \int_0^t {\textbf H}(t-t') \cdot \delta {\textbf a }(t') dt' +{\textbf f}(t),
\label{ddeltaadtnonmarkov}
\end{equation}
where ${\textbf H}(t-t') $ is a  $N\times N$ matrix of memory functions, and  with the additive noise ${\textbf f}(t)$ assumed not  necessarily Gaussian nor $\delta$-correlated, \emph{but  necessarily stationary}, with zero mean ($\langle{\textbf f}(t)\rangle^{ss}={\textbf 0}$), uncorrelated with the initial condition $\delta {\textbf a }(0)$ ($\langle{\textbf f}(t)\delta {\textbf a }^{\dagger}(0)\rangle^{ss}=0$), and with a two-times correlation function given by
\begin{equation}
\langle{\bf f}(t){\bf f}^{\dagger}(t')\rangle^{ss}=  {\bm{\Gamma}}(t-t').
\label{brownnoisecorrfunct}
\end{equation}

Then, the \emph{stationarity theorem}  \cite{delrio} states that stationarity alone is a necessary and sufficient \emph{mathematical} condition for Eq. (\ref{ddeltaadtnonmarkov}) to be written in a very stringent and rigid format, that we shall refer to as the generalized Langevin equation (GLE), namely, 
\begin{equation}
\frac{d\delta  {\textbf a }(t)}{dt}= - {\bm{\omega}} \cdot {\bm{\sigma}}^{ss-1}
\cdot \delta  {\textbf a }(t) - \int_0^{t}dt'  {\bm{\Gamma}}(t-t')\cdot
{\bm{\sigma}}^{ss-1}\cdot \delta {\textbf a }(t')+{\textbf f }(t),
\label{gle01}
\end{equation}
with ${\bm{\omega}}$ being an antisymmetric matrix (${\bm{\omega}}=-{\bm{\omega}}^{\dagger}$), the matrix $ {\bm{\Gamma}}(t)$ having the symmetry $ {\bm{\Gamma}}(t) =  {\bm{\Gamma}}^{\dagger}(-t)$ (which follows from its definition in Eq. (\ref{brownnoisecorrfunct})), and with the matrix  ${\bm{\sigma}}^{ss}$ being the stationary covariance ${\bm{\sigma}}^{ss} \equiv \langle\delta {\textbf a}(t)\delta {\textbf a}(t)^{\dagger}\rangle ^{ss}$. By definition, ${\bm{\sigma}}^{ss}$ is an $N\times N$ symmetric matrix, so that only $N(N-1)/2$ of its $N^2$ elements are independent. The symmetry properties of the matrices ${\bm{\omega}}$ and ${\bm{\Gamma}}(t)$,  imposed by the stationarity condition, imply similar selection rules on the elements of these matrices, which substantially reduces the number of independent elements. In addition, other selection rules may be imposed by other physical symmetry requirements. For example, let us highlight that, if the variables $ a_{i}(t)$ have a definite parity upon time reversal, $a_{i}(-t)=\lambda_i a_{i}(t)$ with  $ \lambda_i =$ 1 or -1, then $\sigma_{ij}^{ss}=\lambda_i\lambda_j \sigma _{ij}^{ss}$, \ \ $\omega _{ij}=- \lambda_i\lambda_j\omega _{ij}$, and $L_{ij}(t)=\lambda_i\lambda_j L _{ij}(t)$ \cite{delrio}.



Eq. (\ref{gle01}) constitutes the mathematical core of an important and well-known statistical mechanical formalism, referred to as the generalized Langevin equation (GLE). Conventionally, Eq. (\ref{gle01}) is  derived using Mori-Zwanzig projection operator  techniques \cite{kubo,zwanzig,mori}  (see Ref. \cite{hansen} for a textbook  presentation, or Sect. 2.2 of Ref. \cite{janssen} for a more  concise account). In essence, the macroscopic variables grouped in ${\textbf a }(t)$ are ultimately dynamical variables in the sense of classical mechanics \cite{goldstein}, i.e., they depend on time through their functional dependence on the phase-space vector $\chi$ of coordinates and momenta of all the constituent particles, i.e., ${\textbf a }(t)={\textbf a }[{\chi(t)}]$. The time-evolution of  ${\textbf a }(t)$ may be formally written as  ${\textbf a }(t)= e^{i\mathcal{L}t}{\textbf a }(0)$, where $\mathcal{L}$ is the so-called Liouvillian operator, which thus governs the full dynamics of our variables of interest. This exact expression, in turn, can be formally projected into the set of ``slow'' variables and its orthogonal part by means of a projection operator, thus allowing us to rewrite the time-evolution equation of the slow variables precisely as  Eq. (\ref{gle01}).

The mathematical structure of this stochastic equation, however,  derives solely from the condition of stationarity \cite{delrio},  and hence, is not a consequence of the Hamiltonian basis of the Mori-Zwanzig projector operator method, which assumes at the outset thermodynamic equilibrium conditions. In fact, the mathematical model represented by the stochastic  equation in Eq. \eqref{gle01} is the route \cite{delrio} to incorporate memory effects in the Onsager and Machlup theory of thermal fluctuations \cite{onsagermachlup1,onsagermachlup2}, which then becomes a phenomenological version of Mori-Zwanzig's mechanistic theory of thermal fluctuations at \emph{equilibrium}. The main advantage of conceiving Eq. (\ref{gle01}) simply as a mathematical model of a stationary stochastic process, is that it can be used to describe properties of systems in stationary states, and not necessarily \emph{thermodynamic equilibrium states}, which are always stationary but must also satisfy the physical condition of thermodynamic equilibrium (i.e. absence of fluxes or maximum entropy). 


\subsection{Coupled tracer and collective diffusion}

Eq. (\ref{gle01}) was first employed in its phenomenological Onsager version in Ref. \cite{faraday} to describe tracer diffusion phenomena in an equilibrium colloidal suspension (a simpler account can also be consulted in Appendix B of Ref. \cite{todos1}). In this theoretical discussion, a relevant sub-product  was derived (see Eq. (4.3) of \cite{faraday}), namely, the WL relation in Eq. (\ref{wertheimlovett1}). In what follows we revisit this derivation, but this time we leave out the assumption that the Brownian colloidal fluid is in thermodynamic equilibrium. Instead, we shall have in mind a stationary but non-equilibrium Brownian system (such as some vibrated granular materials \cite{eliz_grains} or homogeneously stirred suspensions \cite{OUgranular}) to highlight the arguments and steps that demonstrate that, in reality, the WL equation (\ref{wertheimlovett1}) is also valid under more general  stationary non-equilibrium conditions.

\begin{figure*}[ht]
\centering
\includegraphics[width=8cm]{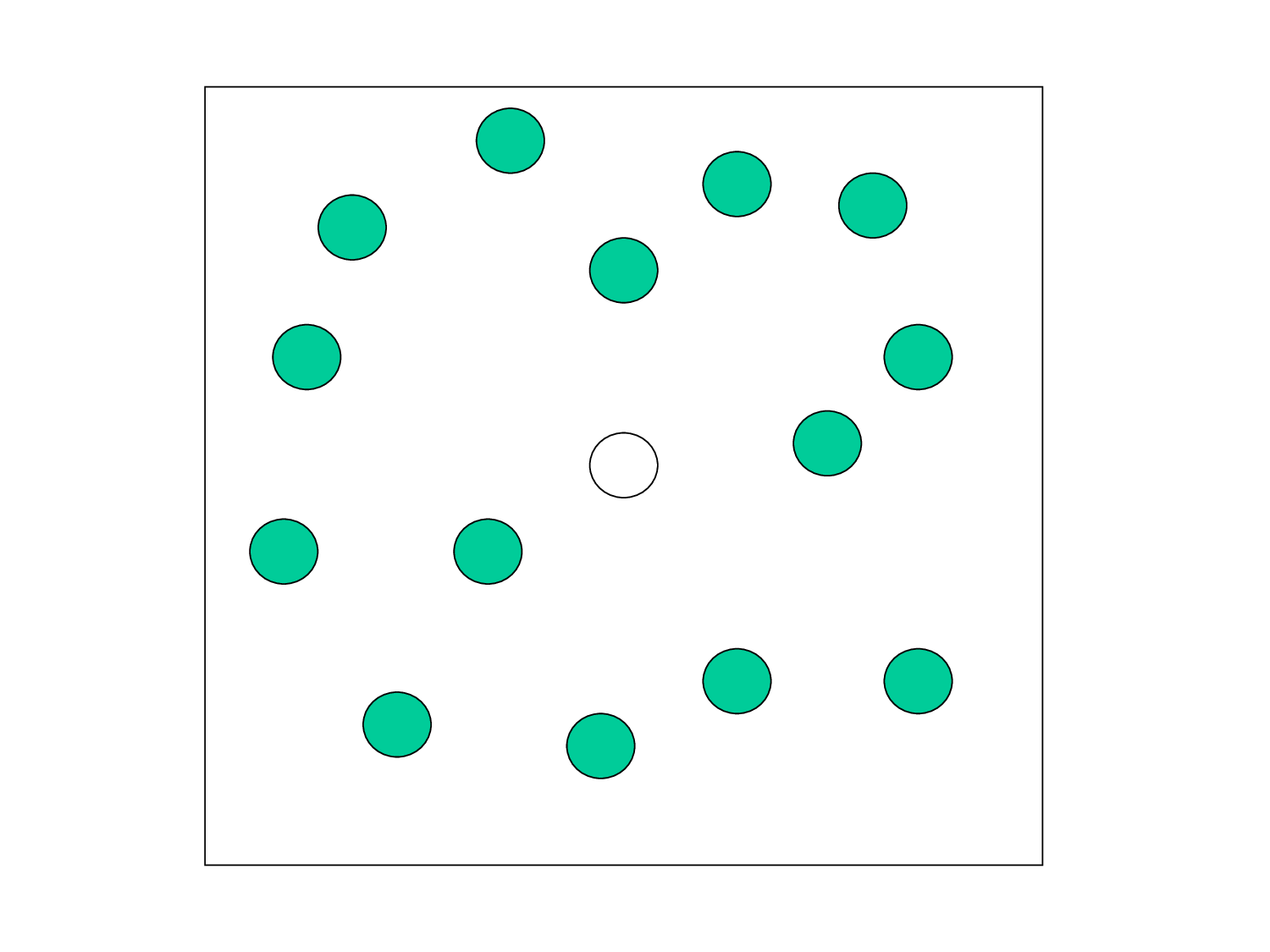}
\caption{A labelled tracer Brownian particle (empty circle) interacting with the surrounding particles (filled circles) of a stationary Brownian fluid.}
\label{tracer}
\end{figure*}

In self-diffusion experiments, the Brownian motion of a very small fraction of labeled particles is recorded, and each of these tracer particles may be regarded as diffusing independently of each other, while interacting with all the un-labelled particles of the suspension (except for the labelling, we assume that the tracer and the host particles are identical). Thus, the state of this system may be represented by a statistical ensemble of identical systems, each containing $N$ identical particles plus a single tracer particle in a volume $V$ (see schematic representation in Fig. \ref{tracer}). Let the state of this system be described by the velocity ${\bf V}(t)$ of the tracer particle, and by the local concentration 
\begin{equation}
\hat{n}'({\bf r},t)\equiv \sum _{i=1}^N\delta (\mathbf{r}-\mathbf{r}_i(t)) 
\label{ndrtmicro}
\end{equation} 
of the surrounding host colloidal particles. The vector position ${\bf r}$ and the position $\mathbf{r}_i(t)$ of the $i$th particle at time $t$, are referred to the center of the tracer, and the prime superscript is a reminder of this fact.

Under stationary conditions, and in the absence of external fields that cause inhomogeneity and/or  anisotropy, the average value of ${\bf V}(t)$ is $\langle {\bf V}(t)\rangle^{ss} ={\bf O}$ (again, the superscript \emph{ss} refers to a stationary state), and the average of $\hat{n}'({\bf r},t)$, denoted as $n^{ss}(r)\equiv \langle \hat{n}'({\bf r},t)\rangle^{ss}$ (where $\langle \cdot  \cdot \cdot \rangle^{ss}$ means average over an arbitrary stationary ensemble), is just $ng^{ss}(r)$, where $g^{ss}(r)$ is the stationary but non-equilibrium radial distribution function of the Brownian particles around the tracer. Contrary to the equilibrium $g^{eq}(r)$, which in principle can be determined by standard statistical thermodynamic theories \cite{mcquarrie,hansen} given the pair potential $u(r)$, there is no general first-principles theory to determine $g^{ss}(r)$ (despite the fact that this quantity is perfectly measurable \cite{eliz_grains}). Nevertheless, we shall avoid assuming thermodynamic equilibrium, but will continue assuming the physical symmetries of stationarity, spatial homogeneity and isotropy to describe the context of Fig. \ref{tracer}. For this, we choose as state variables the velocity  $\mathbf{V}(t)$ of the tracer particle and the local density of host particles $ \hat{n}'({\bf r},t)$, grouped in the stochastic vector $\mathbf{a}(t)\equiv [\mathbf{V}(t), \hat{n}'({\bf r},t)]$, whose mean value is $  \langle \mathbf{a}(t)\rangle^{ss} = \mathbf{a}^{ss} \equiv [\mathbf{0}, n^{ss}(r)]$. We denote the fluctuations of $\mathbf{a}(t)$ around its mean value $\mathbf{a}^{ss} $ by the stochastic vector $\delta \mathbf{a}(t)\equiv [\mathbf{V}(t), \delta n'({\bf r},t)]$, whose components are $\mathbf{V}(t)$  and $\delta n'({\bf r},t) \equiv  \hat{n}'({\bf r},t)- n^{ss}(r)$. Let us now discuss the consequences of the so-called ``stationarity theorem'' \cite{delrio} on the statistical properties of the stochastic vector $\delta \mathbf{a}(t)$.

Applied to the stochastic vector $\delta \mathbf{a}(t)=[\mathbf{V}(t), \delta n'({\bf r},t)]$, Eq.(\ref{gle01})  implies  that ${\bf V}(t)$ and $\delta n'({\bf r},t)$ satisfy two coupled linear stochastic equations containing the specific physical information of our system, but which must conform to the general mathematical format of Eq. (\ref{gle01}). The first of these two equations is the following Langevin equation for the tracer particle \cite{faraday},
\begin{equation}
M{d{\bf V}(t)\over dt}=-\zeta^S \mathbf{V} (t)+ \mathbf{f}^S(t)+\int
d^3 r[\bigtriangledown u(r)]\delta n'({\bf r},t), 
\label{dvdteqvn}
\end{equation}
whose last term is an {\it exact} mechanical coupling between ${\bf V}(t)$ and $\delta n'({\bf r},t)$. It is just the total force $\int d^3r [\bigtriangledown u(r)] \hat{n}'({\bf r},t)$  instantaneously exerted on the tracer by the other particles distributed according to $\hat{n}'({\bf r},t)$ given by Eq. (\ref{ndrtmicro}). Since $\hat{n}'({\bf r},t)=n^{ss}(r)+\delta n'({\bf r},t)$, and because of the radial symmetry of $n^{ss}(r)$, only the departures $\delta n'({\bf r},t)$ from $n^{ss}(r)$ contribute to this force. The other two terms in eq. (\ref{dvdteqvn}) describe the assumed short-time Brownian motion of the tracer particle and represent the friction forces, which in the absence of hydrodynamic interactions, consists of a dissipative term, $-\zeta^S{\bf V}(t)$, plus a corresponding Gaussian, $\delta$-correlated fluctuating force. Within these assumptions, eq. (\ref{dvdteqvn}) is exact. 

The time-evolution equation for $\delta n'({\bf r},t)$ constitutes the second linear stochastic equation for the vector $[{\bf V}(t),\delta n'({\bf r},t)]$, and has the general form 
\begin{equation} 
{\partial\delta n({\bf r},t)\over\partial t}=[\bigtriangledown
n^{ss}(r)]\cdot {\bf V}(t)-\int^t_{0}dt'\int d^3r'D'({\bf
r},{\bf r'};t- t')\delta n'({\bf r'},t') -\bigtriangledown\cdot {\bf
j'}_{dif} ({\bf r},t).
\label{dndteqvn}
\end{equation}
The term linear in ${\bf V}(t)$ derives from linearizing the exact streaming term, $-\nabla \cdot {\bf
j}_{str}({\bf r},t) = -\nabla \cdot [-\hat{n}({\bf r},t) {\bf V}(t)]$, due to the fact that the vector $\bf r$ in $\hat{n}'({\bf r},t)$ is referred to the center of the tracer (which moves with velocity ${\bf V}(t)$). The memory term in this equation is the most general form of the collective diffusion equation as described from the reference frame of the tracer, and the last term represents the corresponding random fluxes.

Let us use at this point all the selection rules imposed by the applicable physical symmetries of the stochastic vector $\delta \mathbf{a}(t)=[\mathbf{V}(t), \delta n'({\bf r},t)]$, to identify the non-zero elements of the matrices ${\bm{\sigma}}^{ss}$, ${\bm{\omega}}$, and $ {\bm{\Gamma}}(t)$ of the GLE in Eq. (\ref{gle01}). As a result, this matrix equation can be written as the following two sub-matrix equations,
\begin{equation}
\frac{d{\bf V}(t)}{dt}= - {\bm{\omega}}_{Vn} \cdot {\bm{\sigma}}_{nn}^{ss-1}
\cdot \delta  n(t) - \int_0^{t}dt'  {\bm{\Gamma}}_{VV}(t-t')\cdot
{\bm{\sigma}}_{VV}^{ss-1}\cdot  {\textbf V }(t')+{\textbf f }_{V}(t)
\label{gle02}
\end{equation}
and
\begin{equation}
\frac{d\delta n(t)}{dt}= - {\bm{\omega}}_{nV} \cdot {\bm{\sigma}}_{VV}^{ss-1}
\cdot \delta  {\textbf V }(t) - \int_0^{t}dt'  {\bm{\Gamma}}_{nn}(t-t')\cdot
{\bm{\sigma}}_{nn}^{ss-1}\cdot \delta n(t')+{\textbf f }_{n}(t).
\label{gle03}
\end{equation}
These two equations must coincide with Eqs. (\ref{dvdteqvn}) and (\ref{dndteqvn}), respectively. In particular, the term $- {\bm{\omega}}_{Vn} \cdot {\bm{\sigma}}_{nn}^{ss-1} \cdot \delta  n(t)$ of Eq. (\ref{gle02}) must coincide with the mechanical term $+\int d^3r[\bigtriangledown u(r)]\delta n'({\bf r},t)$ of Eq. (\ref{dvdteqvn}), whereas the term $- {\bm{\omega}}_{nV} \cdot {\bm{\sigma}}_{VV}^{ss-1} \cdot \delta  {\textbf V }(t)$ of Eq. (\ref{gle03}) must coincide with the streaming term $[\bigtriangledown n^{eq}(r)]\cdot {\bf V}(t)$ of Eq. (\ref{dndteqvn}). 

Since we know that ${\bm{\sigma}}_{nn}^{ss}({\bf r},{\bf r'})= \sigma^{ss} ({\bf r},{\bf r'})$, and that ${\bm{\sigma}}_{VV}^{ss}=(\hat{\beta}M){\bf I}$, it is now not difficult to show that the antisymmetry condition ${\bm{\omega}}_{Vn}^\dagger=-{\bm{\omega}}_{nV}$ can be written as 
\begin{equation}
\bigtriangledown n^{ss}(\mathbf{r})=-\hat{\beta}\int d^3r'\sigma^{ss} ({\bf r},{\bf r'})\nabla 'u(\mathbf{r}'), 
\label{stationaryWLrelation}
\end{equation} 
which is the non-equilibrium (i.e. stationary) extension of the Wertheim-Lovett relation in Eq. (\ref{wertheimlovett1}). In this equation,  $\sigma^{ss} ({\bf r},{\bf r'}) \equiv \langle \delta n'({\bf r},t)  \delta n'({\bf r}',t) \rangle ^{ss}$ is the non-equilibrium stationary covariance. 
Eq. (\ref{stationaryWLrelation}) is, thus, an exact result, involving the stationary covariance $\sigma^{ss} ({\bf r},{\bf r'})$ 
This quantity, however, is in reality not a two-particle but a three-particle correlation function, since it is defined in the presence of the tracer particle (centered at the origin of the vectors ${\bf r}$ and ${\bf r'}$). In what follows, we shall neglect the effects of the external field of the tracer particle on $\sigma^{ss} ({\bf r},{\bf r'})$, and approximate this function by its isotropic and homogeneous version, $\sigma^{ss} ({\bf r},{\bf r'})\approx \sigma^{ss} (\mid \textbf{r}-\textbf{r}'\mid)$. Within these restrictions, the local density $n^{ss}(\textbf{r})$ can be written as $n^{ss}(r)= ng^{ss}(r)$, so that $[ \nabla n^{ss}(r)]= n\nabla[1+ h^{ss}(r)]$, with $g^{ss}(r)=1+h^{ss}(r)$ being the non-equilibrium radial distribution function. Within the same restrictions, the FT $\sigma^{ss}(k)=nS^{ss}(k)$ of $\sigma^{ss}(r)$ is essentially the non-equilibrium SF, $S^{ss}(k)$, and the Wertheim-Lovett relation can be written, in  Fourier space, as 
\begin{eqnarray}
[ i\mathbf{k} h^{ss}(k)]= - \hat{\beta} S^{ss}(k)[ i\mathbf{k} u(k)].
\label{stationaryWLrelation2}
\end{eqnarray}
with $u(k)$ being the Fourier transform of $u(r)$.

In this manner, we see that the WL equation above, originally derived as an exact equilibrium relation \cite{evans}, is in reality a consequence of a more general condition, namely, stationarity. During aging, of course, a glass-forming liquid is not stationary. However, a fundamental assumption of the NE-SCGLE theory is that the real non-equilibrium relaxation of such a liquid can be described approximately as a piece-wise stationary process \cite{nescgle1,nonlinonsmach}, i.e., as a sequence of infinitesimally stationary intervals. Within this approximation, Eq. \eqref{stationaryWLrelation2} will be valid at any stage of the globally non-stationary process, i.e., at any waiting time $t$. Thus, the above WL equation holds replacing $h^{ss}(k)$ and $S^{ss}(k)$ by $h(k;t)$ and $S(k;t)$, respectively. As discussed in the following section, this equation will find a precise use in the derivation of a closed expression for the non-equilibrium shear stress relaxation function $\eta(\tau;t)$ of a glass-forming  colloidal liquid.

\section{Non-equilibrium linear viscoelasticity of a colloidal liquid}\label{section4}

As mentioned in the introduction, one of the main original contributions of this work is the construction of a theoretical scheme for the description of the non-equilibrium linear viscoelastic properties of a fluid after a sudden quench into a glass (or gel) state. For clarity, it is instructive to start by reviewing some pertinent definitions and general relations in the context of the linear viscoelastic response of a colloidal liquid in the absence of hydrodynamic interactions \cite{naegele,banchio}.

\subsection{Linear Viscoelasticity: General relations}\label{subsection3.1}

Let us consider a colloidal suspension of $N$ identical spherical particles with diameter $\sigma$ in a volume $V$, interacting through a radially-symmetric pairwise potential $u(r)$ and subjected to the action of a weak oscillatory shear flow of frequency $\omega$ and shear rate amplitude $\dot{\gamma}_0$. The fluid flow velocity is assumed to be given by the real part of  $\mathbf{u}(\mathbf{r},\tau)=\dot{\gamma}_0y\hat{\mathbf{x}}e^{i\omega \tau}$, where $\hat{\mathbf{x}}$ is the unit vector in the $x$-direction. For simplicity, let us only consider the limit of sufficiently small shear rate amplitudes, $\dot{\gamma}_0 \to 0$, in which the isotropic and homogeneous structure of the suspension is not significantly distorted. In this limit, the linear relationship ${\bm{\Sigma}}(t)=\int_{0}^{t}dt'\eta(t,t')\mathbf{E}(t')$ between the macroscopic stress tensor ${\bm{\Sigma}}(t)$ and the rate of strain tensor $\mathbf{E}(t)$ constitutes the phenomenological definition of the  isotropic and homogeneous total shear stress relaxation function $\eta(t,t')$. Under stationary conditions, ${\bm{\Sigma}}(t)$ and $\mathbf{E}(t)$ become the constants ${\bm{\Sigma}}$ and $\mathbf{E}$, and $\eta(t,t')$ becomes $\eta(t,t')=\eta(t-t')$, so that the linear relationship above becomes ${\bm{\Sigma}}= \eta \mathbf{E}$, with the constant  $\eta$ being the ordinary macroscopic viscosity $\eta\equiv \int_0^\infty \eta(\tau)d\tau$.

In what follows, however, we shall consider our homogeneous suspension to be initially in thermal equilibrium during the time  $-\infty < t \le 0$ at an initial density $n_i=N/V$ and temperature $T_i$. This system is then  subjected at $t=0$ to an instantaneous quench in control parameters to the final values $n$ and $T$. In response, the system  must adjust itself over the time $t>0$ to new stationary conditions. During this relaxation transient, the non-stationary total shear stress relaxation function $\eta(t,t')$ can be written as $\eta(t,t')=\eta(t-t';t)$, i.e., as $\eta(\tau;t)$, with $\tau\equiv t-t'$. This is the non-equilibrium  shear stress relaxation function referred to in the introduction, whose Fourier-Laplace transform $\eta(\omega;t)$  is the dynamic shear viscosity, related with the dynamic shear modulus $G(\omega;t)$ by  $G(\omega;t) =  i\omega \eta(\omega;t)$, whose real and imaginary parts are the elastic and loss moduli $G'(\omega;t)$ and $G''(\omega;t)$.


The total shear stress relaxation function, $\eta(\tau;t)$, can be written as $\eta (\tau;t)= 2\delta (\tau) \eta^0 + \Delta \eta (\tau;t)$, with $\eta^0$ being the ``short-time'' (or ``infinite-frequency'') viscosity, related with the ``short-time'' (or ``free'') self-diffusion coefficient  $D^0$ by the Stokes-Einstein relation $\eta^0= k_BT/3\pi\sigma D^0$. The function $\Delta\eta (\tau;t)$ is the contribution to $\eta (\tau;t)$ due to the inter-particle forces. In the absence of hydrodynamic interactions, $\eta^0$ is the viscosity of the pure solvent, but under some circumstances, such as for concentrated hard-sphere suspensions, the effects of hydrodynamic interactions act virtually instantaneously, simply renormalizing the value of $\eta^0$ and $D^0$, but otherwise behaving as if hydrodynamic interactions were absent \cite{prlhi,mazurgeigen}. This will be a general assumption in what follows.

Our main purpose now is to obtain an approximate but general expression for the function $\Delta \eta (\tau;t)$ in terms of both, the non-equilibrium structure factor $S(k;t)$ and the $t$-evolving and $\tau$-dependent intermediate scattering function $F(k,\tau;t)$. For this, our starting point is the Green-Kubo relation \cite{hansen,boonyip} that can be obtained from the fluctuation-dissipation relation, namely
\begin{equation}
\Delta \eta (\tau;t)= (\beta/V)\langle \sigma ^{xy}(t+\tau) \sigma ^{xy}(t)\rangle, 
\label{greenkubo0}
\end{equation}
where $\sigma ^{xy}(t)$ is the microscopic expression for the configurational component of  the stress tensor, given by \cite{naegele,boonyip}
\begin{equation}\label{eq: sigma1}
\sigma^{xy}(t)  = -\hat{\mathbf{x}}\cdot \left[ \sum_{i=1}^N\mathbf{r}_i(t)\mathbf{F}_i(t)\right] \cdot\hat{\mathbf{y}}
\end{equation}
where $\mathbf{r}_i(t)$ and $\mathbf{F}_i(t)$ are the position and total force on the $i$-th colloidal particle, respectively. As before, in Eq. (\ref{greenkubo0}), the brackets $\langle ... \rangle$ indicate a general (not necessarily  \emph{equilibrium}) ensemble average. 

Also, in the same equation, $\beta\equiv 1/k_BT$, where $T$ is the molecular temperature \textcolor{blue}{$T(t)$} defined in Eq. (\ref{molectemp}), assumed to coincide with the final temperature of the quench. This, however, entails another drastic simplification that must be made explicit here. We refer to the assumption that the system is in contact with a thermal reservoir at temperature $T^R(t)$, and that heat is conducted instantaneously through the surface and bulk of the fluid, so that the local time-dependent molecular temperature $T(t)$ defined in Eq. (\ref{molectemp}) as  $T(t)\equiv   \langle \mathbf{p}^2(t)/3Mk_B\rangle$, is uniform and equal to $T^R(t)$. For the particular case of an instantaneous temperature quench at $t=0$ from an initial temperature $T_i$  to a final  temperature $T$ that remains constant for $t>0$, the temperatures $T(t)=T^R(t)$ will remain constant, or $T(t)=T^R(t)=T$ for  $t>0$. 

In the absence of external fields, 
we may rewrite equation \eqref{eq: sigma1} as
\begin{equation}
 \sigma ^{xy}(t) = -\sum_{i=1}^N R_i^x(t) F_i^y(t) = \frac{1}{2}\sum_{ i, j =1}^N x_{ij}\frac{du(R_{ij})}{dy_{ij}},
\label{defsigmaxy}
\end{equation}
where $R_i^x(t) \equiv \hat{\mathbf{x}}\cdot \mathbf{r}_i(t)=x_i(t)$, $F_i^y(t)\equiv \mathbf{F}_i(t) \cdot\hat{\mathbf{y}}$, $x_{ij}(t)=x_i(t)-x_j(t)$, and ${du(R_{ij})/dy_{ij}}\equiv (\nabla_{ij} u(R_{ij})) \cdot \hat{\mathbf{y}}$. In terms of the local density of particles, 
one can rewrite Eq. \eqref{defsigmaxy} as 
\begin{eqnarray}
 \sigma ^{xy}(t)&=&\frac{1}{2}\int d\mathbf{r} \int d\mathbf{r}' (x-x') \frac{du(\mid \mathbf{r}-\mathbf{r}'\mid)}{d(y-y')} \sum_{i=1}^N \delta(\mathbf{r}-\mathbf{r}_i(t))\sum_{ j=1}^N \delta(\mathbf{r}'-\mathbf{r}_j(t))\nonumber \\
&=&\frac{1}{2}\int d\mathbf{r} \int d\mathbf{r}' (x-x') \frac{du(\mid \mathbf{r}-\mathbf{r}'\mid)}{d(y-y')} \hat{n}(\mathbf{r},t)\hat{n}(\mathbf{r}',t), 
\label{defsigmaxy1} 
\end{eqnarray}
which, inserted  in Eq. (\ref{greenkubo0}), leads to 
\begin{eqnarray}
\Delta \eta (\tau;t)&=& \frac{\beta}{4V}\int d\mathbf{r}_1  d\mathbf{r}_2 d\mathbf{r}_3 d\mathbf{r}_4(x_1-x_2) \frac{du(\mid \mathbf{r}_1-\mathbf{r}_2\mid)}{d(y_1-y_2)} (x_3-x_4) \frac{du(\mid \mathbf{r}_3-\mathbf{r}_4\mid)}{d(y_3-y_4)}  \nonumber \\
& & \times \Big\langle \hat{n}(\mathbf{r}_1,t+\tau)\hat{n}(\mathbf{r}_2,t+\tau)\hat{n}(\mathbf{r}_3,t)\hat{n}(\mathbf{r}_4,t)\Big\rangle.
 \label{deltaeta1}
\end{eqnarray}

Using  the Fourier transform $u(k) = \int d\mathbf{r} e^{-i\mathbf{k}\cdot \mathbf{r}} u(r)$ of the pair potential $u(r)$, it is straightforward to show that $-\mathbf{r}\nabla u(r) = (1/(2\pi)^3)\int d\mathbf{k} e^{i\mathbf{k}\cdot \mathbf{r}} \nabla_\mathbf{k}[\mathbf{k}u(k)]$, whose $xy$ component (writing $\mathbf{r}=\mathbf{r}_1-\mathbf{r}_2$) is
\begin{equation}
(x_1-x_2) \frac{du(\mid \mathbf{r}_1-\mathbf{r}_2\mid)}{d(y_1-y_2)} = \frac{-1}{(2\pi)^3}\int d\mathbf{k} e^{i\mathbf{k}\cdot ( \mathbf{r}_1-\mathbf{r}_2)} \hat{\mathbf{x}} \cdot \nabla_\mathbf{k}[\mathbf{k} \cdot \hat{\mathbf{y}} u(k)] = \frac{-1}{(2\pi)^3}\int d\mathbf{k} e^{i\mathbf{k}\cdot ( \mathbf{r}_1-\mathbf{r}_2)} \left( \frac{\partial \left[k_yu(k)\right]}{\partial k_x} \right).
\label{greenkubo}
\end{equation}
This result allows us to rewrite Eq. (\ref{deltaeta1}) as 
\begin{eqnarray}
\Delta \eta (\tau;t)&=& \frac{(\beta/V)}{4(2\pi)^6}\int d\mathbf{k}\int d\mathbf{k}' \left( \frac{\partial \left[k_yu(k)\right]}{\partial k_x} \right)\left( \frac{\partial \left[k'_yu(k')\right]}{\partial k'_x} \right)\int d\mathbf{r}_1  d\mathbf{r}_2 d\mathbf{r}_3 d\mathbf{r}_4 e^{i\mathbf{k}\cdot ( \mathbf{r}_1-\mathbf{r}_2)}e^{i\mathbf{k}'\cdot ( \mathbf{r}_3-\mathbf{r}_4)} \nonumber  \\
& &\times \Big\langle \hat{n}(\mathbf{r}_1,t+\tau)\hat{n}(\mathbf{r}_2,t+\tau)\hat{n}(\mathbf{r}_3,t)\hat{n}(\mathbf{r}_4,t)\Big\rangle, 
\label{deltaeta2}
\end{eqnarray}
which can also be written as 
\begin{eqnarray}
\Delta \eta (\tau;t)&=& \frac{(\beta/V)}{4(2\pi)^6}\int d\mathbf{k}\int d\mathbf{k}' \left( \frac{\partial \left[k_yu(k)\right]}{\partial k_x} \right)\left( \frac{\partial \left[k'_yu(k')\right]}{\partial k'_x} \right) \Big\langle n(\mathbf{k},t+\tau)n(-\mathbf{k},t+\tau)n(\mathbf{k}',t)n(-\mathbf{k}',t)\Big\rangle, \nonumber  \\
\label{deltaeta21}
\end{eqnarray}
where $n(\mathbf{k},t)$ is the Fourier transform $n(\mathbf{k},t)\equiv \int d\mathbf{r}\ e^{i\mathbf{k}\cdot \mathbf{r}} n(\mathbf{r},t)$ of $n(\mathbf{r},t)$.

\subsection{Expression for $\eta(\tau;t)$ in terms of $S(k; t)$ and $F(k,\tau; t)$}\label{maineq}

Eq. \eqref{deltaeta21} writes $\Delta \eta (\tau;t)$ in terms of  the Fourier-transform $\langle n(\mathbf{k},t+\tau)n(-\mathbf{k},t+\tau)n(\mathbf{k}',t)n(-\mathbf{k}',t)\rangle$ of the  four-point correlation function $\langle \hat{n}(\mathbf{r}_1,t+\tau)\hat{n}(\mathbf{r}_2,t+\tau)\ \hat{n}(\mathbf{r}_3,t)\hat{n}(\mathbf{r}_4,t)\rangle$, whose calculation is probably impossible without some form of simplifying approximation. For this, here we adopt its Gaussian factorization, which approximates $\Big\langle n(\mathbf{k},t+\tau)n(-\mathbf{k},t+\tau)n(\mathbf{k}',t)n(-\mathbf{k}',t)\Big\rangle$ by a sum of products of two-point and one-point correlation functions. In Appendix \ref{AppendixC} we demonstrate that, restricting ourselves to the case of a homogeneous and isotropic liquid, such factorization allows us to write Eq. (\ref{deltaeta21}) as 
\begin{eqnarray}
\Delta \eta (\tau;t)&=& \frac{\beta n^2}{2(2\pi)^3}\int d\mathbf{k}\int d\mathbf{k}' \left( \frac{\partial \left[k_yu(k)\right]}{\partial k_x} \right)\left( \frac{\partial \left[k'_yu(k')\right]}{\partial k'_x} \right) F(k,\tau;t)F(k',\tau;t)\delta(\mathbf{k}+\mathbf{k}').\nonumber  \\ 
\label{deltaetaApp4}
\end{eqnarray}

At this point, let us  substitute in this equation, the expression for $[k_z u(k)]$ provided by the WL relation derived in the previous section, Eq. \eqref{stationaryWLrelation2}, namely, $[k_y u(k)]= - k_BT S^{-1}(k;t)[k_y h(k;t)]$. This allows us to rewrite Eq. \eqref{deltaetaApp4} as
\begin{eqnarray}
\label{deltaeta4}
\Delta \eta (\tau;t)&=& 
 \frac{\beta n^2}{2(2\pi)^3}\int d\mathbf{k}  \left( \frac{\partial \left[k_yu(k)\right]}{\partial k_x} \right)^2   \left[ F(k,\tau;t) \right] ^2
\nonumber \\ &=&   \frac{k_BT n^2}{2(2\pi)^3}\int d\mathbf{k}  \left( \frac{\partial \left[k_yS^{-1}(k;t) h(k;t)\right]}{\partial k_x} \right)^2   \left[ F(k,\tau;t) \right] ^2
\nonumber \\ &=& \frac{k_BT}{2(2\pi)^3}\int d\mathbf{k} k_y^2 \left( \frac{\partial \left[1-S^{-1}(k;t)\right]}{\partial k_x} \right)^2   \left[ F(k,\tau;t) \right] ^2,
\end{eqnarray}
which can be more conveniently rewritten as 
\begin{eqnarray}
\label{deltaeta4}
\Delta \eta (\tau;t)&=&    \frac{k_BT}{2(2\pi)^3}\int d\mathbf{k} \left(\frac{ k_xk_z}{k} \right)^2 \left[ \frac{1}{S(k;t)} \left(\frac{d S(k;t)}{d k} \right) \right]^2   \left[ \frac{F(k,\tau;t)}{S(k;t)} \right] ^2.
\end{eqnarray}
Upon angular integration $\int d\mathbf{k} \left(\frac{ k_xk_z}{k} \right)^2 f(k)= \left( \frac{4\pi}{15} \right) \int_0^\infty k^4 dk f(k)$, this expression reads 
\begin{eqnarray}
\label{deltaeta4}
\Delta \eta (\tau;t)&=& \frac{k_BT}{60\pi^2}  \int_0^\infty dk k^4   \left[ \frac{1}{S(k;t)} \left(\frac{d S(k;t)}{d k} \right) \right]^2   \left[ \frac{F(k,\tau;t)}{S(k;t)} \right] ^2,
\end{eqnarray}
which is the expression for $\eta(\tau;t)$ in terms of $S(k; t)$ and $F(k,\tau; t)$ that we set out to derive. Let us finally highlight that the structure of this equation is identical to that derived for thermodynamic equilibrium conditions by Geszti  for atomic fluids \cite{geszti} and  by  N\"agele and Bergenholtz \cite{naegele,banchio} using mode coupling theory.

\section{Discussion and summary}\label{section5}

This work was aimed to contribute to the discussion of the relevant and general challenge of extending fundamental concepts of the statistical mechanical theory of classical equilibrium liquids, to out-of-equilibrium conditions. In addressing this challenge our specific motivation and perspective derived from the development of the statistical physical formalism referred to as the \emph{non-equilibrium self-consistent generalized Langevin equation} theory of irreversible processes of liquids  \cite{nescgle0,nescgle1,nescgle2,nescgle3,nescgle5}. Given its successful first-principles description of aging and other essential fingerprints of the amorphous solidification of liquids, we deemed important to highlight some methodological aspects implicit in the derivation of the NE-SCGLE equations, since they will continue to be employed in further extensions and applications of this non-equilibrium theoretical approach. 

The simplest of these methodological aspects consisted in revising the derivation of a given set of theoretical results of the equilibrium theory of liquids, to see if they really employ the actual condition of thermodynamic equilibrium, through the use, for example, of an equilibrium (canonical, microcanonical, ...) probability distribution function. As it happens, many steps in these equilibrium derivations actually employ only the condition of stationarity (but not of thermodynamic equilibrium), as well as other temporal or spatial symmetries (spatial homogeneity and/or isotropy, time-reversal, etc.). Our method was first illustrated in Section \ref{section2} with a simple exercise, namely, the derivation of the non-equilibrium energy equation, followed in Section \ref{section3} by a second example,  the derivation of the non-equilibrium extension of the Wertheim-Lovett relation. 

These two specific illustrative examples clearly prepared the stage for the main specific contribution of this work, namely, the derivation of the non-equilibrium extension of an expression -- first  derived by Geszti  \cite{geszti} and  by  N\"agele and Bergenholtz \cite{naegele,banchio}--  for the rheological and viscoelastic properties of liquids in terms of the structural and dynamical properties of the system. This extension, carried out in Section \ref{section4}, led us to an approximate but general expression, Eq. \eqref{deltaeta4}, that connects the non-equilibrium  shear stress relaxation function $\eta(\tau;t)$ of a non-equilibrium liquid  with the kinetics of the structural relaxation, encoded in the $t$-evolution of the non-equilibrium structure factor $S(k;t)$, and the dynamic correlations represented by the (collective and self) intermediate scattering functions $F(k,\tau;t)$ and  $F_S(k,\tau;t)$. These are, according to Eq. \eqref{deltaeta4}, the main microscopic elements that determine the value of  $\eta(\tau;t)$ and of the instantaneous viscosity $\eta(t)\equiv \int_0^{\infty}  d \tau \eta(\tau;t)$, thus directly relating the viscoleastic response of a glass- or gel- forming system with explicit microscopic details, such as the potential of interaction between the constitutive particles, and the protocol of fabrication (here simplified by considering only an instantaneous quench). To the best of our knowledge, such a connection had never been established before.

As we shall demonstrate in a separate work, Eq. (\ref{deltaeta4}), together with the NE-SCGLE equations Eqs. (\ref{dsktenst})-(\ref{fluctsquench}), constitute a proposal of a canonical theoretical protocol to determine the viscoelasticity of non-equilibrium liquids from first-principles. The resulting approach is now ready for its systematic application to the characterization of the viscoelastic response of a diversity of qualitatively different glass and gel forming  systems, such as liquids with Lennard Jones-like interactions \cite{olais1,olais2} or systems with competing interactions (short-ranged attraction plus long-ranged repulsion) \cite{carretas}. In these systems, the interference between thermodynamical instabilities (spinodal line, $\lambda$-line) and dynamical arrest mechanisms leads to the possibility of qualitatively different glassy states, ranging from porous glasses, gels and Wigner glasses \cite{wigner,camacho}.

Although rather secondary to the main line of arguments just described, in Section \ref{section2} we also  addressed the natural question of the non-equilibrium extension of the Ornstein-Zernike equation $S^{eq}(k)=1/n\mathcal{E}(k;n,T)$. There, we concluded that this equation is actually a condition for thermodynamic equilibrium, and that the deviations $\left[S(k;t) -1/n\mathcal{E}(k;n,T)\right]$, according to Eq. (\ref{dsktenst}), drive the rate of change of the structure of the liquid (represented by $S(k;t)$). 

Let us finally notice that there are no fundamental barriers that prevent the extension of the arguments and equations presented here, to much more complex conditions, involving glass and gel forming systems with multiple relaxation channels. This is the case, for example, of colloidal suspensions comprised by dipolar particles (ferrofluids), in which the decoupling of the orientational and translational dynamics allows to investigate partially arrested states, and also, of mixtures with disparate size ratios, which allow for the development of glassy states with qualitatively different structural and dynamical characteristics upon tuning the molar distribution and total concentration. The discussion of the non-equilibrium viscoelastic response of these more complex materials is an additional example of areas of opportunity left for subsequent work.

\medskip
\section{Acknowledgments}
ACKNOWLEDGMENTS: This work was supported by the Consejo Nacional de Ciencia y Tecnolog\'ia (CONACYT, Mexico) through Postdoctoral Fellowships Grants No. I1200/224/2021 and I1200/320/2022; and trough grants 320983, CB A1-S-22362, and LANIMFE 314881.

\vskip3cm
\appendix

\section{Gaussian approximation for the four-point correlation function and Fourier transforms}\label{AppendixC}


This Appendix discusses the Gaussian factorization for the four-point correlation function $\Big\langle n(\mathbf{k},t+\tau)n(-\mathbf{k},t+\tau)n(\mathbf{k}',t)n(-\mathbf{k}',t)\Big\rangle$ under more general conditions \cite{isserlis,dhont} than employed in the MCT description of  equilibrium viscoelasticity \cite{naegele}. For clarity, let us introduce the following notation for the four microscopic densities involved, namely, $n(\mathbf{k},t+\tau)\equiv n_1$, $n(-\mathbf{k},t+\tau)\equiv n_2$, $n(\mathbf{k}',t)\equiv n_3$, and $n(-\mathbf{k}',t)\equiv n_4$; and for their averages and fluctuations, $\langle n_i\rangle\equiv \overline n_i$ and $\delta n(\mathbf{r}_i;t)\equiv \delta n_i$ (with $i=1, 2, 3$, and 4). Then, 

\begin{eqnarray}
    & &\Big\langle n(\mathbf{k},t+\tau)n(-\mathbf{k},t+\tau)n(\mathbf{k}',t)n(-\mathbf{k}',t)\Big\rangle \nonumber\\
    &=&\langle n_1 n_2 n_3 n_4 \rangle=\langle(\bar{n}_1+\delta n_1)(\bar{n}_2+\delta n_2)(\bar{n}_3+\delta n_3)(\bar{n}_4+\delta n_4)\rangle\nonumber\\ 
    &=& \langle\delta n_1\delta n_2\delta n_3\delta n_4\rangle + \langle\delta n_1\delta n_2\delta n_3\rangle\bar{n}_4 + \langle\delta n_1\delta n_2\delta n_4\rangle\bar{n}_3 + \langle\delta n_1\delta n_3\delta n_4\rangle\bar{n}_2\nonumber\\
    & & + \langle\delta n_2\delta n_3\delta n_4\rangle\bar{n}_1 +\langle\delta n_1\delta n_3\rangle\bar{n}_2\bar{n}_4 + \langle\delta n_1\delta n_2\rangle\bar{n}_3\bar{n}_4 +\langle\delta n_2\delta n_3\rangle\bar{n}_1\bar{n}_4\nonumber\\
    & & +\langle\delta n_3\delta n_4\rangle\bar{n}_1\bar{n}_2 +\langle\delta n_1\delta n_4\rangle\bar{n}_2\bar{n}_3 + \langle\delta n_2\delta n_4\rangle\bar{n}_1\bar{n}_3 +\langle\delta n_1\rangle\bar{n}_2\bar{n}_3\bar{n}_4\nonumber\\
    & & +\langle\delta n_2\rangle\bar{n}_1\bar{n}_3\bar{n}_4+\langle\delta n_3\rangle\bar{n}_1\bar{n}_2\bar{n}_4+\langle\delta n_4\rangle\bar{n}_1\bar{n}_2\bar{n}_3+\bar{n}_1\bar{n}_2\bar{n}_3\bar{n}_4.
\end{eqnarray}

If each of the variables $n_i$ above represented a stationary Gaussian stochastic process, then from Isserlis-Wick's theorem \cite{isserlis,dhont} it would follow that 
\begin{eqnarray}
    \langle\delta n_i\rangle &=& \langle\delta n_i\delta n_j\delta n_k\rangle = 0,\\
    \langle\delta n_1\delta n_2\delta n_3\delta n_4\rangle &=& \langle\delta n_1\delta n_2\rangle \langle\delta n_3\delta n_4\rangle + \langle\delta n_1\delta n_3\rangle \langle\delta n_2\delta n_4\rangle + \langle\delta n_1\delta n_4\rangle \langle\delta n_2\delta n_3\rangle,
\end{eqnarray}
The stochastic process represented by the variables $n_i$ above is not strictly stationary. However, it will be  assumed to be  piece-wise stationary \cite{nescgle1}. It is also not necessarily Gaussian. Nevertheless, we adopt this factorization as an approximation. As a result, and after some straightforward algebraic steps, one gets
\begin{eqnarray}
    \left\langle n_1n_2n_3n_4\right\rangle\nonumber &\approx&
   (\bar{n}_1\bar{n}_2+\langle\delta n_1\delta n_2\rangle)(\bar{n}_3\bar{n}_4+\langle\delta n_3\delta n_4\rangle) + (\bar{n}_1\bar{n}_3+\langle\delta n_1\delta n_3\rangle)(\bar{n}_2\bar{n}_4+\langle\delta n_2\delta n_4\rangle)  \nonumber \\ 
& & + (\bar{n}_1\bar{n}_4+\langle\delta n_1\delta n_4\rangle)(\bar{n}_2\bar{n}_3+\langle\delta n_2\delta n_3\rangle) - 2\bar{n}_1\bar{n}_2\bar{n}_3\bar{n}_4 \nonumber \\
&=& \langle n_1 n_ 2\rangle\langle n_3 n_ 4\rangle+ \langle n_1 n_ 3\rangle\langle n_2 n_4\rangle
+\langle n_1 n_4\rangle\langle n_2 n_3\rangle- 2\bar{n}_1\bar{n}_2\bar{n}_3\bar{n}_4.\label{ggaa}
\end{eqnarray}
Going back to the original notation, this equation reads
\begin{eqnarray}
\Big\langle n(\mathbf{k},t+\tau)n(-\mathbf{k},t+\tau)n(\mathbf{k}',t)n(-\mathbf{k}',t)\Big\rangle &\approx& \Big\langle n(\mathbf{k},t+\tau)n(-\mathbf{k},t+\tau)\Big\rangle  \Big\langle  n(\mathbf{k}',t)n(-\mathbf{k}',t)\Big\rangle
\nonumber \\
& & + \Big\langle n(\mathbf{k},t+\tau) n(\mathbf{k}',t)\Big\rangle  \Big\langle n(-\mathbf{k},t+\tau)n(-\mathbf{k}',t)\Big\rangle \nonumber \\ & &
 + \Big\langle n(\mathbf{k},t+\tau)n(-\mathbf{k}',t)\Big\rangle  \Big\langle  n(\mathbf{k}',t)n(-\mathbf{k},t+\tau)\Big\rangle \nonumber \\
& & - 2\langle n(\mathbf{k},t+\tau)\rangle \langle n(-\mathbf{k},t+\tau)\rangle \langle n(\mathbf{k}',t)\rangle \langle n(-\mathbf{k}',t)\rangle \nonumber \\
\label{deltaetaApp0}
\end{eqnarray}

By translational invariance, the van Hove function $G(\mathbf{r},t;\mathbf{r'},t')\equiv\langle n(\mathbf{r};t)n(\mathbf{r}';t')\rangle$ can only depend on the difference $\mathbf{r}-\mathbf{r'}$ and by spatial isotropy,  $G(\mathbf{r}-\mathbf{r'};t,t')$ can only depend on the magnitude  $\mid \mathbf{r}-\mathbf{r'}\mid$. Similarly, as a consequence of translational invariance, the correlation $\langle n(\mathbf{k},t) n(\mathbf{k}',t')\rangle$ is non-zero only if $\mathbf{k}'=-\mathbf{k}$ and by rotational invariance, $F(\mathbf{k}; t,t')$ can only depend on the magnitude of $\mathbf{k}$, i.e., $N^{-1}\langle n(\mathbf{k},t) n(\mathbf{k}',t')\rangle=F(\mathbf{k},t;\mathbf{k}',t')= F(k; t,t')\ \delta(\mathbf{k}+\mathbf{k}')$. On the other hand, the mean values $\langle n(\mathbf{r},t) \rangle$ and $\langle n(\mathbf{k},t) \rangle$ depend in general on $t$, but in the present application they are constrained for simplicity to be uniform and constant, $\langle n(\mathbf{r},t) \rangle = n \equiv N/V$, so that $\langle n(\mathbf{k},t) \rangle = (N/V)(2\pi)^3 \delta (\mathbf{k})$. The two-time correlation function $F(k; t,t')$, which under stationary conditions only depends on the time difference $\tau\equiv t-t'$, depends in general on both times, $t$ and $t'$ (or, equivalently, on $t$ and $\tau\equiv t-t'$), so that we shall actually write $N^{-1}\langle n(\mathbf{k},t) n(\mathbf{k}',t')\rangle= F(k,\tau; t)\ \delta(\mathbf{k}+\mathbf{k}')$. Finally, let us notice that the equal-time intermediate scattering function $F(k,\tau=0; t)$ is just the time-dependent structure factor $S(k;t)$, i.e.,    $F(k,\tau=0; t)=S(k;t)$. 

With these previsions, Eq. (\ref{deltaetaApp0}) becomes 
\begin{eqnarray}
\Big\langle n(\mathbf{k},t+\tau)n(-\mathbf{k},t+\tau)n(\mathbf{k}',t)n(-\mathbf{k}',t)\Big\rangle &\approx&  
S(k;t+\tau)S(k';t) 
\nonumber \\
& &
+N^{2}F(k,\tau;t)F(k',\tau;t)(2\pi)^3V^{-1} \delta(\mathbf{k}+\mathbf{k}') \nonumber \\
& & +N^{2}F(k,\tau;t)F(k',\tau;t) (2\pi)^3V^{-1}\delta(\mathbf{k}-\mathbf{k}')  \nonumber \\
& &
- 2(n/V)^2(2\pi)^6  \delta (\mathbf{k}) \delta (\mathbf{k}'),  \nonumber  \\
\label{deltaetaApp1}
\end{eqnarray}
which allows us to approximate $\Delta \eta (\tau;t)$ in Eq. (\ref{deltaeta21}) by
\begin{eqnarray}
\Delta \eta (\tau;t)&=& \frac{(\beta/V)}{4(2\pi)^6}\int d\mathbf{k}\int d\mathbf{k}' \left( \frac{\partial \left[k_zu(k)\right]}{\partial k_x} \right)\left( \frac{\partial \left[k'_zu(k')\right]}{\partial k'_x} \right) \nonumber  \\
& & \times \Big\lbrace S(k;t+\tau)S(k't)+N^2F(k,\tau;t)F(k',\tau;t)(2\pi)^3V^{-1}\delta(\mathbf{k}+\mathbf{k}')  \nonumber  \\
& & +N^2F(k,\tau;t)F(k',\tau;t)(2\pi)^3V^{-1}\delta(\mathbf{k}-\mathbf{k}')-2\left(\frac{n}{V}\right)^2(2\pi)^6 \delta(\mathbf{k})\delta(\mathbf{k}')\Big\rbrace. 
\label{deltaetaApp2}
\end{eqnarray}
Thus, $\Delta \eta (\tau;t)$ is a sum of four terms. The first of them becomes a product of two factors, each vanishing because the integrand is an odd function of $k_z$. The last term also vanishes because it also factorizes, with each factor being proportional to   $\int d\mathbf{k} \ k_z(\partial u(k)/\partial k_x)\ \delta(\mathbf{k})= [k_z(\partial u(k)/\partial k_x)]_{\mathbf{k}=\mathbf{0}}=0$. Thus, the only surviving terms are the second and the third, which are clearly identical. These considerations allow us to rewrite Eq. (\ref{deltaetaApp2}) as 
\begin{eqnarray}
\Delta \eta (\tau;t)&=& \frac{\beta n^2}{4(2\pi)^3}\int d\mathbf{k}\int d\mathbf{k}' \left( \frac{\partial \left[k_zu(k)\right]}{\partial k_x} \right)\left( \frac{\partial \left[k'_zu(k')\right]}{\partial k'_x} \right) \nonumber  \\
& & \times \Big\lbrace 2F(k,\tau;t)F(k',\tau;t)\delta(\mathbf{k}+\mathbf{k}') \Big\rbrace, 
\label{deltaetaApp3}
\end{eqnarray}
which is Eq. (\ref{deltaetaApp4}).

\eject

\vskip5cm


\end{document}